\documentclass[10pt,a4paper,notitlepage]{revtex4-1} 

\setcitestyle{numbers}

\usepackage{subfig}

\usepackage[utf8]{inputenc}

\usepackage[english]{babel}

\usepackage{xcolor}

\usepackage{afterpage}		
\usepackage{graphicx}
\usepackage{latexsym}
\usepackage{amsmath}
\usepackage{amssymb}

\renewcommand{\appendix}{
\setcounter{section}{0}
\renewcommand{\thesection}{\Roman{section}}
\vspace{0.5cm}
{\Large{\bf APPENDIX}}}

\newtheorem{theorem}{Theorem}[section]

\newtheorem{postulate}[theorem]{Postulate}

\newtheorem{remark}[theorem]{Remark}

\let\origbar\bar
\let\origdot\dot
\let\bar\origbar
\let\dot\origdot

\newcommand{\Ss}{\mathbb{S}}

\newcommand{\Rbb}{\mathbb{R}}

\newcommand{\nn}{\bf n}
\newcommand{\mm}{\bf m}
\newcommand{\uu}{{\bf u}}
\newcommand{\xb}{{\bf x}}

\newcommand{\ww}{{\bf w}}
\newcommand{\vv}{{\bf v}}
\newcommand{\rr}{{\bf r}}
\newcommand{\fb}{{\bf f}}
\newcommand{\gb}{{\bf g}}
\newcommand{\hb}{{\bf h}}
\newcommand{\pb}{{\bf p}}

\newcommand{\Fv}{{\bf F}}
\newcommand{\Gv}{{\bf G}}
\newcommand{\Hv}{{\bf H}}

\newcommand{\lbf}{{\bf l}}

\newcommand{\qb}{{\bf q}}

\begin{document}

\title[Mass diffusion]{A new Eulerian model for viscous and heat conducting compressible flows}
\author{Magnus Sv\"ard}
\affiliation{Department of Mathematics\\ 
University of Bergen \\
Postbox 7800 \\
5020 Bergen \\
Norway }
       \email{Magnus.Svard@uib.no}

\date{\today}

\begin{abstract}
In this article, a suite of physically inconsistent properties of the Navier-Stokes equations, associated with the lack of mass diffusion and the definition of velocity, are presented. We show that these inconsistencies are consequences of  the Lagrangian derivation that models viscous stresses rather than diffusion.

A new model for compressible and diffusive (viscous and heat conducting) flows  of an ideal gas, is derived in a purely Eulerian framework. We propose that these equations supersede the Navier-Stokes equations.

A few numerical experiments demonstrate  some differences and similarities between the new system and the Navier-Stokes equations.

\end{abstract}

\maketitle

\section{Introduction}



The Navier-Stokes(-Fourier) equations that model a perfect, compressible, viscous and heat conducting  gas at local thermodynamic equilibrium, in three space-dimensions are,
\begin{align}
\partial_t \rho + div_\xb(\rho \vv )&= 0, \nonumber \\
\partial_t (\rho \vv) + div_\xb(\rho \vv \otimes \vv) + \nabla_\xb p&=div_\xb \Ss, \label{NS} \\
\partial_t (E) + div_\xb(E \vv +p\vv)  &=div_\xb \Ss\vv +div_\xb(\kappa \nabla_{\xb}T ),\nonumber\\
p&=\rho R T, \quad \textrm{ideal gas law,}\nonumber  
\end{align}
where $\rho,\rho \vv, E$ are the conserved variables density, momentum and total energy. $\xb=(x,y,z)$ are Cartesian coordinates. $\vv=(u,v,w)^T$ is the velocity vector. $p,T$ are pressure and temperature; $\Ss$ is the stress tensor (for a Newtonian fluid) and  $\mu$ and $\kappa$ are
the viscosity and thermal diffusivity coefficients.  The total energy is given by $E=\frac{p}{\gamma-1}+\frac{\rho |\vv|^2}{2}$ where $\gamma=c_p/c_v$ and $c_{p,v}$ are the heat capacities at constant pressure or volume. Furthermore, $R$ is the gas constant. Moreover, when setting the right-hand side of (\ref{NS}) to zero, the equations are termed the Euler equations and model an inviscid, non-diffusive fluid.
\begin{remark}
The stress tensor generally contains two viscosity coefficients. For brevity, we use a standard approximation of the second coefficient, $\lambda=-\frac{2}{3}\mu$ (see \cite{Tritton}).   
\end{remark}
\begin{remark}
Herein, we do not consider non-equilibrium thermodynamics or any intermolecular forces.
  \end{remark}

The equations (\ref{NS}) are widely accepted as an accurate flow model and their importance in engineering and physics can not be understated. Solutions to them are routinely approximated using Computational Fluid Dynamics (CFD).  

Despite their wide acceptance, they have been challenged. Recently, by H. Brenner (see \cite{Brenner05_1,Brenner05_2,Brenner13, BardowOttinger07,GreenshieldsReese07,GuoXu09}) who argued that the velocity appearing in the inviscid terms and in the viscous terms are not equal. The two differ by a mass diffusive flux, he argued. (Other alternative models to the Navier-Stokes system that include diffusion in the continuity equation are found in \cite{Goddard10,DadzieReese08,DadzieReese10,DadzieReese12,DadzieReese12_2}.) 

Brenner's modification changed the  mathematical structure of (\ref{NS}), which is hyperbolic in the continuity equation and parabolic in the others, to become (completely) parabolic. This allowed for a proof of  weak well-posedness of a system related to Brenner's (see \cite{FeireislVasseur09}) and in \cite{Svard15} a convergent numerical scheme was derived. For (\ref{NS}) no such results are known. In fact, no results of well-posedness, neither existence, uniqueness, nor continuity of data, is available for the system (\ref{NS}) and for the incompressible version it has been suggested that the system is not well-posed \cite{Tao16}. However, Brenner's system was rebutted on physical grounds in \cite{Ottinger09}.

The problem of the unknown well-posedness of (\ref{NS}) is not merely a mathematical nuisance. It hampers the design of effective CFD methods. Deriving a reliable, robust, and convergent numerical scheme requires knowledge of the mathematical function spaces that the solutions occupy.  It may be tempting to attribute the lack of well-posedness results to the inability of mathematicians (see e.g. \cite{Svard16}). Nevertheless, it should be uncontroversial that well-posedness with respect to  existence and uniqueness of solutions is a necessary feature for reliable predictions.  (Continuity with respect to data is generally desirable but might be locally be violated due to non-linear instabilities.)

However, this paper is not about the well-posedness of the Navier-Stokes system. The motivation for this article is: 1) to examine some physical properties of the Navier-Stokes equations; 2) to study the physical assumptions on which the Navier-Stokes equations are based; 3) to propose a new system based on the previous insights. 

The paper is organized as follows: In Section \ref{sec:critical}, a number of physically inconsistent properties of the Navier-Stokes equations are presented. Many, or all, of these problems are well-known and they are linked to the definition of velocity and the non-diffusive mass flux. Based on the discussion of multi-component flows,  a new continuity equation that includes mass diffusion is derived in Section \ref{sec:new_cont}. However, the arguments pointing to a mass diffusive continuity equation do not reveal how to modify the remaining equations of the Navier-Stokes system. One option is to augment (\ref{NS}) with a mass diffusive contribution that obey an entropy principle (see \cite{GuermondPopov14}). However, such models were rebutted in \cite{Ottinger09}. 

Therefore, the derivation of (\ref{NS}) is studied in Section \ref{sec:stress}. This investigation is largely decoupled from the observations in Section \ref{sec:critical} and leads to the root of the problems with the Navier-Stokes equations, which is an incompatible mix of arguments using either Lagrangian mass parcels or Eulerian control volumes, and the modelling of viscous stresses rather than diffusion.

To avoid inconsistencies, a purely Eulerian framework is used, and conservation properties and symmetries are established in Section \ref{sec:av_var}. (The approach to derive a continuum model in the Eulerian frame has previously been used in  \cite{Goddard10}.)  A novel Eulerian model is derived in Section \ref{sec:new_model} and we propose that it supersedes (\ref{NS}). The new model includes mass diffusion and resolves the issues presented in Section \ref{sec:critical}. 

In Section \ref{sec:numerics}, the new Eulerian model and the Navier-Stokes equations are solved numerically and the results compared.

\begin{remark}
The purpose of deriving a new model is \emph{not} to extend the range of applicability beyond that of the standard  Navier-Stokes(-Fourier) model. Nor is it to ``fix'' the equations to become mathematically appealing. The purpose is to obtain a physically consistent model.
\end{remark}

\section{Unphysical properties of the Navier-Stokes equations}\label{sec:critical}

\subsection{Positivity}\label{subsec:pos}

 Positivity of density and temperature is  a well-known physical constraint. Since solutions of (\ref{NS}) have to be obtained by some limiting process from finite dimensional approximations, we exemplify its behaviour in a numerical context. A common reason for catastrophic failure of numerical codes is that density becomes negative. (E.g. at strong bow shocks.) This can only be prevented by adding artificial diffusion, which implies that the following continuity equation is effectively solved (in 1-D),
\begin{align}
\rho_t+(\rho u)_x=ch\rho_{xx},\label{mod_eqn}
\end{align}
where $h$ is a measure of the grid size and $c$ a coefficient proportional to $|u|$. Clearly, the right-hand side will disappear as $h\rightarrow 0$ but its presence ``guides'' the limiting process and ensures that $\rho>0$. The equations (\ref{NS}) themselves do not have this mechanism although their parabolic terms (i.e., the right-hand side) are in principal much stronger than the artificial diffusion in  (\ref{mod_eqn}). (They do not vanish as $h\rightarrow 0$.) 

However, it is undesirable to add artificial diffusion since it severely decreases the numerical accuracy and it is not Galilean invariant. The fact that it disappears as $h\rightarrow 0$ may still not be enough to avoid frame dependency between inertial systems.

On the other hand, if mass diffusion is added as a parabolic term in the equations, it is possible to ensure positivity, by proving that $\ln \rho$ or $\rho^{-1}$ are bounded. (See \cite{FeireislVasseur09,Svard15}.) Notably, the mass diffusive coefficient need not be proportional to $|u|$, which rids the problem  from the lack of Galilean invariance. (See Section \ref{sec:galilean}.)

\subsection{Adiabatic Wall}\label{sec:adiabatic1}

Beginning   with the non-diffusive Euler equations (for which ``adiabatic'' carries no meaning), only one boundary condition is necessary at a wall:  $\vv\cdot{\nn}=0$ (no-penetration). It implies that there is no flow of mass, momentum and energy through a wall, which is physically reasonable. Furthermore, the equations can be recast into a hyperbolic equation for the entropy, $S=\ln(p/\rho^\gamma)$. In 1-D, it is
\begin{align}
(-\rho S)_t+(-\rho uS )_x=0.\label{entEuler}
\end{align} 
\begin{remark}
There are several ways to phrase an entropy equation. This is preferable from a mathematical perspective since an upper bound on $-\rho S$ implies a bound on the conservative variables in $L^2$, see e.g. \cite{Dafermos,Svard15}. 
\end{remark}
Integrating (\ref{entEuler}) over the domain $(0,1)$ yields, 
\begin{align}
\int_0^1(-\rho S)_t\,dx=\rho u S|_0^1.
\end{align}
With the normal component of velocity being zero at a wall, the boundary terms vanish. Hence, there is no entropy flow through a wall. In the Euler model, the governing equations,  the entropy equation, and the boundary conditions consistently model ``no-flow'' through a wall for an inviscid fluid.

Next, we turn to the Navier-Stokes equations. At an adiabatic wall, the system (\ref{NS}) has to be augmented with boundary conditions. For an adiabatic wall, they are $\vv=0$ (no-slip) and  $\frac{\partial T}{\partial n}=0$. (See e.g. \cite{SvardNordstrom08} for a linear well-posedness analysis.) In 1-D, the entropy equation is
\begin{align}
(-\rho S)_t+(-\rho uS )_x=-\kappa (\ln T)_{xx}-\frac{4\mu}{3}\frac{u_x^2}{T}-\kappa \frac{T_x^2}{T^2}.\label{NSent}
\end{align}
Equation (\ref{NSent}) describes the transport and \emph{diffusion} of entropy.  
Integrating (\ref{NSent}) over $(0,1)$ yields the same convective boundary terms as in the Euler case, which vanish thanks to the boundary condition $u=0$. In addition, the first term on the right-hand side yields a boundary term $\kappa T_x/T$. The adiabatic boundary condition, $T_x=0$, ensures that it vanishes and an upper bound on $-\rho S$ is obtained. (The two last terms are negative.) However, the normal derivative of the entropy is,
\begin{align}
\frac{\partial S}{\partial x}=(1-\gamma)\frac{\partial \rho}{\partial x}+\frac{\partial T}{\partial x}.\label{ent_norm_der}
\end{align}
which is non-zero at a wall since $\rho_x$ is in general non-zero. Usually, the gradient of a variable is associated with a diffusive flux, and hence there appears to be a non-zero entropy diffusion through the adiabatic wall despite the fact that (\ref{NSent}) models diffusive effects; $S_x\neq 0$ is inconsistent with a diffusive model. (This observation is due to \cite{Mark}.)

Furthermore, at an adiabatic wall, the physically correct behaviour is that no mass or energy flow through the wall. In the Navier-Stokes equations, (\ref{NS}), $\rho_x\neq 0$ may be interpreted as a diffusive flux through the wall. (However, it may be argued that (\ref{NS}) does not model mass diffusion and therefore $\rho_x\neq 0$  is not associated with a diffusive flux.)   

Moreover, the energy equation is undeniably a diffusive model. Using the gas law and the boundary conditions $u=0,T_x=0$, we obtain,
\begin{align}
E_x=\left(\frac{p}{\gamma-1}+\frac{1}{2}\rho u^2 \right)_x = \frac{R}{\gamma-1}(\rho_x T+\rho T_x)+\frac{1}{2}\rho_xu^2+\rho uu_x=
\frac{R}{\gamma-1}(\rho_x T)
\end{align}
Clearly, $\rho_x\neq 0$ leads to an energy diffusion through the adiabatic wall. 
\begin{remark}
The momentum need not be zero since there is in general a transfer of momentum from objects to the flow. The momentum diffusion is
\begin{align}
(\rho u)_x=\rho u_x+\rho_xu \nonumber 
\end{align}
which is not zero in general even if $\rho_x=0$.
\end{remark}
We have pointed to three unphysical properties of the Navier-Stokes equations: there is diffusion of mass, total energy and entropy through an adiabatic wall. 

Note that, if mass diffusion is included in the model, an extra boundary condition is needed, $\rho_x=0$ (or generally the normal derivative, $\partial_n \rho=0$), which implies that $\rho_x=S_x=E_x=0$. (See Section \ref{sec:wall_eul}.) 

\begin{remark}
Note that the Navier-Stokes equations \emph{are} conservative with respect to mass, energy and entropy with standard wall boundary conditions. What we claim is that the modelling is unphysical. In diffusive models the normal gradient usually signifies a diffusive flux through the boundary. In (\ref{NS}), the normal gradient can not be interpreted as a diffusive flux.
\end{remark}

\subsection{Far-field boundaries}

The analysis of boundary conditions is strongly linked to the question of well-posedness of the system. Since a complete non-linear well-posedness theory is lacking, it is common to resort to linear well-posedness, which is a necessary, albeit not sufficient, requirement. 

An important part of linear well-posedness on bounded domains is the analysis of boundary conditions. Not only should they model the correct physics, mathematically they should be the smallest set that results in a bounded solution.  If too many boundary conditions are imposed the solution may become multi-valued on the boundary, inducing discontinuities in the solution. Too few boundary conditions imply that some variables are undefined. (Both cases nullify the physics as well.)  However, even if the correct number of boundary conditions are imposed, it may still not lead to a bounded solution.  The boundary conditions must be able to continuously change character  according to the flow conditions, or else discontinuities and/or instabilities may be induced.

To analyze (linear) well-posedness, the system (\ref{NS}) is linearized, symmetrized, and the coefficients are frozen. (See \cite{KreissLorenz,SvardCarpenter07}). We obtain a system system of the form,
\begin{align}
\ww_t+A\ww_x+B\ww_y+C\ww_z=(\Fv_V)_x+(\Gv_V)_y+(\Hv_V)_z,\label{linear_symm}
\end{align}
\begin{align}
\Fv_V&=B_{11}\ww_x+B_{12}\ww_y+B_{13}\ww_z,\nonumber \\
\Gv_V&=B_{21}\ww_x+B_{22}\ww_y+B_{23}\ww_z,\nonumber \\
\Hv_V&=B_{31}\ww_x+B_{32}\ww_y+B_{33}\ww_z,\nonumber 
\end{align}
where $A,B,C$ are symmetric matrices and the block matrix $[B_{ij}]$ is symmetric \emph{positive
semi-definite} for (\ref{NS}). (See e.g. \cite{NordstromSvard05} and \cite{KreissLorenz}.) Since there is no mass diffusion in (\ref{NS}), the first row of all $B_{ij}$ is 0. Furthermore, $\ww$ denotes the vector of symmetrized variables.

Far-field boundary conditions are used far away from objects immersed in a flow. The assumption is that close to the boundary, the flow is hardly affected by the object such that the flow is a free stream. 

To simplify notation, we assume a square domain, $\Omega=\{0<x,y,z<1\}$. Consider the inviscid Euler equations, where $B_{ij}=0$. In this case, the system is purely hyperbolic and the most common (well-posed) far-field boundary conditions specify the in-going characteristics,
\begin{align}
A^+\ww=A^+\gb(t), \quad x=0.\label{bc_euler}
\end{align}
Here,  $A^+=X\Lambda^+X^T$ where $X$ is the eigenvector matrix and $\Lambda$ the (diagonal) eigenvalue matrix. $\Lambda$ is divided by the sign of its eigenvalues as $\Lambda=\Lambda^++\Lambda^-$. $\gb(t)$ is a vector of known and bounded functions. (Typically the free stream values of the solution vector $\ww$.) 

Although the boundary condition is stated for the linearized system it is straightforward to translate it to the non-linear setting of the original equations. The main advantage with this boundary condition is that it automatically sets the correct number of boundary conditions at every point individually on the boundary. For example, this allows a point to be a subsonic inflow while its neighbour is a subsonic outflow. (This scenario is common when swirls, vortices, or turbulence in viscous flows, approach the boundary. )

Turning to the Navier-Stokes equations, it is common to use a mixed boundary condition based on (\ref{bc_euler}).
\begin{align}
A^+\ww-\Fv_V&=A^+\gb(t), \quad x=0.\label{bc_NS}
\end{align}
If the flow is approximately a free-stream, i.e., $\Fv_V\approx 0$, this boundary condition should essentially behave as (\ref{bc_euler}). The condition (\ref{bc_NS}) implies that: 5 boundary conditions are specified on sub- and supersonic inflows; 5 boundary conditions on subsonic outflows; 4 boundary conditions on  supersonic outflows. However, at a subsonic outflow only 4 boundary conditions should be specified. (See \cite{NordstromSvard05}.) Hence, at a subsonic outflow, (\ref{bc_NS}) over-specifies  (\ref{linear_symm}),
which becomes ill-posed. The remedy is to exchange  $A^+$ with a matrix $A'$ in the outstanding case, 
\begin{align}
A'\ww-\Fv_V&=A'\gb(t), \quad x=0, \quad \textrm(subsonic\,\, outflow)\label{bc_NS2}
\end{align}
such that four conditions are imposed and the solution remains bounded.  (See \cite{SvardCarpenter07} for the complete analysis and a definition of
$A'$.) This procedure makes the linear (constant coefficient) system well-posed but there are  (at least) two unfavourable consequences.
\begin{itemize}
\item If the velocity is fluctuating around 0, the non-linear version of the boundary
  condition has to change locally between (\ref{bc_NS}) and
  (\ref{bc_NS2}). This is a non-smooth procedure and it is questionable if the
  resulting solution will be smooth (or even stable). 
\item For any flow with subsonic outflow boundaries, the Navier-Stokes system (\ref{NS}) is
  incompatible with the Euler system. (If ${\bf F}_V$ vanishes, we are left with $A'\ww=A'\gb$, not (\ref{bc_euler}), which over-specifies the Euler system.) \emph{Hence, the Euler system can not be the inviscid limit of the Navier-Stokes equations.}

\end{itemize}
If the Navier-Stokes system included mass diffusion, the system becomes completely parabolic and an extra boundary condition is needed. For such a system, five boundary conditions should always be used, irrespective of the Mach number and the flow direction. Hence, (\ref{bc_NS})  works in all cases and  both disadvantages are resolved. (Obviously, $\Fv_V$ must be augmented with the mass flux terms and becomes a full vector. See also Section \ref{sec:farf_eul}.)

\subsection{Relaxation to thermodynamic equilibrium}\label{sec:relaxNS}

A gas consists of distinct molecules that travel at different speeds and collide with each other. The point values of the macroscopic field variables used in the Euler or Navier-Stokes equations represent  averages of the particle properties in an equilibrated neighbourhood. The neighbourhood over which the average is taken has to be large enough to contain sufficiently many molecules but also small enough such that collisions between molecules quickly, much faster than macroscopic time scales, even out differences within the control volume.  That is,  the fluid, governed by (\ref{NS}), is in \emph{local thermodynamic equilibrium.}  We call the scale of such neighbourhoods the \emph{continuum scale}.

The macroscopic variables can in principle not be determined below the continuum scale but since the model is a Partial Differential Equation (PDE) they are formally defined at an arbitrarily short length scale. Therefore, one must choose a behaviour of the PDE at short length scales. The natural behaviour, consistent with the assumption of local thermodynamic equilibrium, is that diffusion is dominating at sub-continuum scales.

To study this mechanism,  we will for brevity  consider the 1-D model on a bounded periodic domain  $\Omega=-L\leq x \leq L$.  In 1-D, (\ref{NS}) reduces to,
\begin{align}
\rho_t + (m)_x&=0 \nonumber \\
m_t + (um+p)_x&=(\frac{4\mu}{3} u_x)_x \label{NS1D}\\
E_t + (u(E+p))_x&=(\frac{4\mu}{3} u u_x + \kappa T_x)_x, \nonumber\\
p&=\rho R T,\quad \textrm{ideal gas law},\nonumber 
\end{align}
where $m=\rho u$ is the momentum.

Consider a fluid at rest, $u=0$, and that $m=0$, at $t=0$ with no pressure gradient, $p_x=0$. From the ideal gas law we have,
\begin{align}
0&=p_x=\rho_x RT + \rho RT_x,\quad\textrm{or}\quad
0=\frac{p_x}{p}=\frac{\rho_x}{\rho}+ \frac{T_x}{T}.\nonumber 
\end{align}
Let the density and temperature at time $t=0$, be given as
\begin{align}
\rho&=\exp(-\delta \cos(n 2\pi x/L )),\label{osc} \\
T&=\exp(\delta \cos(n 2\pi x/L )),\nonumber 
\end{align}
such that
\begin{align}
\frac{\rho_x}{\rho}=-\frac{T_x}{T}=\delta\frac{2n\pi}{L} \sin(n 2\pi x/L ),\quad (p_x=0),\nonumber
\end{align}
where $0<\delta<<1$ is the amplitude.

With these smooth initial data, the Navier-Stokes equations (\ref{NS1D}) predict the following first instant of the flow,
\begin{align}
\rho_t &=0, \nonumber \\
m_t &=0, \nonumber\\
E_t &= (\kappa T_x)_x, \nonumber
\end{align}
at every $x\in \Omega$. Since $\rho_t=m_t=0$, the last equation is
\begin{align}
\frac{1}{\gamma-1}p_t &= (\kappa T_x)_x, \nonumber
\end{align}
at the first instant.  Since $T_x$ is not constant, the temperature  induces an oscillation of the
pressure, $p$, i.e., $p_x$ becomes non-zero. This causes a change of $m$
through the second equation in (\ref{NS1D}), which  in turn affects $\rho$ through the first equation. Most importantly, an oscillatory velocity is induced.

The path to thermodynamic equilibrium from these initial data
is driven by the momentum and energy equations. If the diffusion of
temperature and the viscous stresses are sufficiently strong, the system will (probably) relax to an equilibrium. Throughout this process, the continuity equation redistributes mass but does not contribute to the relaxation, since it has no diffusion term.

Irrespective of how small the wave length of the disturbance (\ref{osc}) is, i.e. how large $n$ is, there is no scale at which the system relaxes without inducing an advective flow.   In the continuity equation, momentum is the sole mechanism for moving mass, effectuated by $u=m/\rho$. Since mass has to be redistributed in order to reach equilibrium, advection is inevitable even at arbitrarily small scales. That is, diffusion is not the dominating process at the sub-continuum scale and \emph{the Navier-Stokes system intrinsically breaks an assumption on which it is based.} 

We remark that, both diffusive processes (temperature and viscosity) are attributed to  particles that randomly travel between regions of different macroscopic states.  They transfer their velocity, causing diffusion of $u$ and $T$, but they do not transfer their mass. By definition there is no diffusive mass transport, or equivalently, diffusive transport is included in $u$.

The discussion leads us to demand that the following principle is satisfied for a diffusive flow model.
\begin{postulate}\label{assump1}
A fluid model should allow a mode of pure (independent of the velocity) thermodynamic relaxation.  
\end{postulate}
In order to satisfy Postulate \ref{assump1}, \emph{mass diffusion} is needed.

\subsection{Multi-component fluid}\label{subsec:two_comp}

When more than one species of fluid is present in a flow, it is well-known that the species will mix, not only by advective transport, but also by diffusion. A multi-component model contains total density, $\rho$, and $N$ partial densities, $\rho_{i}$, one for each fluid. By definition, 
\begin{align}
\sum_{i=1}^N\rho_i(x,t)=\rho(x,t).\label{sum_rho}
\end{align}
The total density satisfies the standard continuity equation in (\ref{NS1D}), i.e.,
\begin{align}
\rho_t + (\rho u)_x&=0. \label{2comp_rho} 
\end{align}
In addition, each component is governed by a continuity equation,
\begin{align}
(\rho_k)_t + (\rho_k(u+j_k))_x&=0, \quad \quad k=\{1,2,...N\}, \label{cont_comp}
\end{align}
where $j_k$ is a diffusive flux. A natural model for the diffusion of a species is
\begin{align}
j_k&=-\tilde D_k (\rho_k)_x, \label{fick_natural} 
\end{align}
where the diffusion coefficient, $\tilde D_k$, may depend on $\rho_k,\rho$ and $T$. However, using (\ref{fick_natural}), implies that equation (\ref{cont_comp}) does not generally reduce to (\ref{2comp_rho}) in the case of a single species. 

Instead, Fick's law that describes diffusion of a species dissolved in a fluid is used. In its standard form it models diffusion of concentration. Interpreting $\rho_k/\rho$ as a density concentration, one obtains the model,
\begin{align}
j_k&=-\frac{\rho}{\rho_k}D_k (\frac{\rho_k}{\rho})_x, \label{fick} 
\end{align}
where $D_k$ is the diffusion coefficient. (Equation (\ref{fick}) is a special case of (\ref{fick_natural}).) Note that for one species, $\rho_1=\rho$, and hence $j_k=0$ and the model reduces to (\ref{2comp_rho}). However, for multiple species, definition (\ref{fick}) makes the system over-specified since the sum of (\ref{cont_comp}) for all species does not equal (\ref{2comp_rho}). Hence, Fick's law has to be further modified 
\begin{align}
j_k&=-\frac{\rho}{\rho_k}D_k (\frac{\rho_k}{\rho})_x+\sum_{j=1}^N D_j \left( \frac{\rho_j}{\rho}\right)_x. \label{fick_cook} 
\end{align}
(See e.g. \cite{Cook09}.) The last sum is added to make  $\sum_{i=1}^N\rho_kj_k=0$, which ensures that (\ref{2comp_rho}) is not violated.

The simplest diffusion model (\ref{fick_natural}), that contains the greatest freedom to model physics by choosing $\tilde D_k$, has to be exchanged for increasingly complicated models ending with (\ref{fick_cook}), in order to not violate (\ref{2comp_rho}).  

On the other hand, if the continuity equation of the total density was simply the sum of the  models for the partial densities, one could return to (\ref{fick_natural}) and choose $\tilde D_k$ based on physical principles and experimental data. Of course, this would introduce mass diffusion in (\ref{2comp_rho}). 

Furthermore, in a multi-component model, the velocity in (\ref{2comp_rho}) is the sole mechanism for transporting the \emph{total density} (including random motions.) However, the same velocity appearing in the partial density equation (\ref{cont_comp}) models pure advection since the diffusive terms model the random motions. Hence, \emph{the velocity does not have a unique meaning and is therefore internally inconsistent for multi-component flows.}

\begin{remark}
We emphasize that (\ref{fick_cook}) is not a unique way to choose a mass diffusive law that satisfies $\sum_{i=1}^N\rho_kj_k=0$. Another example is
\begin{align}
j_k=-D\frac{\rho}{\rho_k}\left(\frac{\rho_k}{\rho}\right)_x\label{fick2}
\end{align}
but it requires $D$ to be a species-independent diffusion constant.
\end{remark}

\subsection{A new continuity equation}\label{sec:new_cont}

Consider a two-component model. ($N=2$ in (\ref{cont_comp}).) If we choose a model for the diffusion of partial densities, rather than a model designed not to violate (\ref{2comp_rho}),  then by sheer symmetry (\ref{cont_comp}) should hold  for both $k=1$ and $k=2$. Moreover, the total mass is by definition given by (\ref{sum_rho}) and the sum of the continuity equations (\ref{cont_comp}) for each component, gives us a new equation that replaces  (\ref{2comp_rho}). 
\begin{align}
(\rho)_t + (\rho u)_x=&-(\rho_1j_1+\rho_2j_2)_x.\label{new_cont}
\end{align}
This continuity equation is now consistent with both component's
continuity equations given by (\ref{cont_comp}).  But most importantly, since the mass diffusion now appears as explicit terms in (\ref{new_cont}), we conclude that \emph{ $u$ represents the   advective velocity (without diffusive transport)} in all three continuity equations. Hence, there is no internal inconsistency regarding the definition of $u$.  

\subsubsection{Two identical fluids}

We proceed with a thought-experiment for (\ref{new_cont}). Consider a single component fluid at rest in a container. That is, there is no advective movement within the fluid. Pressure, temperature and density are all constant. It is clear that no advective velocity will appear in the container if left unperturbed.  

Next, assume now that we randomly ``colour`` (or label) all fluid elements in the container as either red or blue.  The physical properties are the same and the fluid is at rest. It appears reasonable that no advective velocity is created and $\rho$ and $p$ remains constant. It is also expected that the two fluids eventually mix by diffusion.

To analyze this example mathematically, we consider the states  $\rho,T,p$ to be positive constants and $u=u_x=0$. The three continuity equations and the compatibility equation are
\begin{align}
\rho_t+(\rho u)_x&=(-\rho_1 j_1 -\rho_2j_2)_x,\label{two_Fick}\\
(\rho_1)_t+(\rho_1 (u+j_1))_x&=0,\nonumber\\
(\rho_2)_t+(\rho_2 (u+j_2))_x&=0,\nonumber\\
\rho_1+\rho_2&=\rho.\nonumber 
\end{align}
Furthermore, we have $0=\rho_x=(\rho_1)_x+(\rho_2)_x$, or
$(\rho_1)_x=-(\rho_2)_x$. Under the premises that $\rho_t=m_t=0$ for $t>0$, i.e., that
$m,\rho$ remain constant, it follows from (\ref{two_Fick}) that 
\begin{align}
-\rho_1 j_1 -\rho_2j_2=0.\label{prop1}
\end{align}
The simplest possible diffusion that satisfies (\ref{prop1}) is
\begin{align}
j_i=-\nu_i\frac{(\rho_i)_x}{\rho_i}=-\nu_i (\ln \rho_i)_x,\label{new_fick}
\end{align}
where $\nu_i>0$ is a diffusion coefficient. (Since we are considering two
identical fluids, we have $\nu_1=\nu_2$ and (\ref{prop1}) is thereby
satisfied.) 

\begin{remark}
At this point, $\nu>0$ can take a general form, e.g., $\nu=\nu(\rho,T)$.
\end{remark}

By removing the colouring in the previous example, we have two components with the same properties. That is, a single component fluid. Then all three equations (\ref{cont_comp}) and (\ref{new_cont}) collapse to the same equation. 
\begin{align}
\rho_t+(\rho u)_x = (\nu \rho (\ln \rho)_x)_x=(\nu \rho_x)_x.\label{new_cont_single}
\end{align}
Mass diffusive continuity equations, such as (\ref{new_cont_single}), appear to solve all the problems presented in Section \ref{sec:critical}. However, to understand the problems beyond the continuity equation, we take a different view in the coming sections.

\begin{remark}
Equation (\ref{new_cont_single}) is the same as that proposed by Brenner \cite{Brenner05_2} up to the diffusion coefficient $\nu$.  
\end{remark}

\section{The stress tensor}\label{sec:stress}





Equations for inviscid compressible flows are routinely derived by establishing a balance inside a \emph{control  volume} $V$, with sides of length $\delta x=\delta y= \delta z<<1$.  The control volume is fixed in space, i.e., Eulerian, and the fluid passes through it.  The equations are obtained by using three core physical principles: Conservation of mass, momentum and energy. For instance, the mass in the control volume is changing due to the mass flux through the boundary as
\begin{align}
\partial_t \int_V \rho dV +\int_{\partial V} \rho (\vv\cdot {\nn}) \,dS=0,\label{CV_rho}
\end{align}
where $\nn$ is the outward facing normal on the boundary $\partial V$, and $S$ is the infinitesimal surface element. In the inviscid (Euler) approximation diffusive effects are neglected and since there is no mechanism for producing mass there are no other contributions.

Proceeding to the second principle, the average x-momentum inside the volume is  changing due to the transport of x-momentum through the boundary and the pressure  force on the boundary. The balance becomes,
\begin{align}
\partial_t \int_V (\rho u) dV +\int_{\partial V} (\rho u)(\vv\cdot {\nn})
 dS+ \int_{\partial V} p ({\nn}\cdot {{\bf e}}_x) \, dS  =0.\label{momentum_euler}
\end{align}
where ${\bf e}_x$ is the unit vector in the positive x-direction.

Since the equations for density and momentum do not form a closed system, the third principle, energy conservation, must be used. The energy balance in the control volume is obtained by the corresponding arguments. (We omit the details.)

By assuming sufficient regularity the well-known Euler equations are obtained. In order for the resulting system  of partial differential equations to be a valid model, it must support solutions that justify the regularity assumptions made along the way.  In this case, it is well-known that the Euler equations do support discontinuous (shock) solutions and these are obtained by solving the PDE system weakly. That is, by \emph{returning to an integral form} and thereby relaxing the regularity assumptions again.

To add viscous effects, the Lagrangian form is used, which is obtained by turning the control volume into a moving fluid parcel with constant mass.  Hence, the continuity equation is replaced by an equation for the specific volume, $V(\xb,t)$. Assuming that $\vv$ is continuous and that the continuity equation is satisfied strongly, the (Euler) momentum equations can be written as 
\begin{align}
\rho\frac{D\vv}{Dt}=-\nabla_\xb p, \label{NS_mom}
\end{align}
where $D/Dt$ is the material derivative $\partial_t+\vv\cdot \nabla_\xb$.

In (\ref{NS_mom}) the fluid is inviscid and $p$ is the only force acting on the material surface causing acceleration and deformation of the fluid parcel. The Navier-Stokes equations are obtained by augmenting the equations by the viscous forces, modelled by the stress tensor $\Ss $ acting on the parcel surface. The momentum equations become 
\begin{align}
\rho\frac{D\vv}{Dt}=-\nabla_\xb p + \nabla_\xb \Ss. \label{ns_momentum}
\end{align}

To return to the Eulerian frame from the  Lagrangian requires sufficient smoothness of the velocity field (and the other variables) and of $V(\xb,t)$.  \emph{If the equations do not support sufficiently regular solutions, the derivation of the
equations is invalid.} Solving the Navier-Stokes equations in Lagrangian or Eulerian form may thus give different solutions. 

\begin{remark}
Note that well-posedness, generally does not require smoothness. Smoothness is an ``extra'' property of the solution that may, or may not, be possible to prove. Hence, even with a well-posed model at hand, the Lagrangian and Eulerian forms may not be equivalent.
\end{remark}

\begin{remark}
It is not obvious that a fluid parcel, $V(\xb,t)$, would remain smooth in a fully developed turbulent flow, where it stretches, twists, and shears. 
\end{remark}

The situation is to be contrasted with the Euler equations, where smoothness is
required to pose the differential form but one can solve them weakly which removes
the smoothness assumptions.

However, there is also another aspect connected to the Lagrangian view of a fluid parcel of \emph{constant mass}. By definition, it excludes mass diffusion. Moreover, \emph{ a fixed material parcel ought to exclude all diffusion.} However, that is not the case in (\ref{NS}) since  friction and stresses appear due to diffusion. Furthermore, heat diffusion is explicitly modelled  in (\ref{NS}) (via an Eulerian control volume argument). Since diffusion implies that the mass parcel is \emph{not} fixed, the Navier-Stokes model appears to be self-contradictory.

Even more to the point, \emph{in the Navier-Stokes equations the main principle that motivates the parabolic terms is  Newton's Second Law. However, Newton's Second Law is not a first cause or principle in this case. The stress forces appearing in the model are caused by viscosity, which in turn is fundamentally caused by diffusion. (Molecules that transfer their momentum between sheets of fluid due to random motions.) Physically, the stress tensor is a tertiary cause.} 

To model the first cause  (diffusion), an Eulerian frame is natural since diffusion is ruled out in a fixed-mass Lagrangian fluid parcel.

\subsection{Derivation of (\ref{NS}) in Eulerian form}\label{sec:eulerian_ns}

Sometimes the Navier-Stokes equations are derived directly in the
Eulerian framework. The argument is that the stress tensor and pressure exert forces on the control volume boundary. We will take a closer look at the implications of an Eulerian view. 

To simplify the arguments, we consider the two-dimensional case on a square domain. The sides are located at $\{x,y\}=\{0,1\}$. Let us consider the $\rho u$ momentum equation. In the Navier-Stokes model $\rho u$ is evolved by
\begin{align}
(\rho u)_t +(\rho u^2+p)_x+(\rho uv)_y=(\frac{4}{3}\mu u_x +\frac{2}{3}\mu v_y)_x+(\mu u_y+\mu v_x)_y\label{NS_mom1D}
\end{align}
for a Newtonian fluid. Partial integration over $V=[0,1]\times [0,1]$, yields the following boundary terms from the right-hand side of (\ref{NS_mom1D}). 
\begin{align}
\frac{4}{3}\mu u_x +\frac{2}{3}\mu v_y,\quad x=\{0,1\},\label{visc_BT} \\
\mu u_y+\mu v_x,\quad y=\{0,1\}.\nonumber
\end{align}
These boundary terms supposedly model forces on the control volume boundary. Before we analyze them, we recall how stresses appear in a Newtonian fluid. The basic principle is that two fluid sheets that slide against each other generate a friction force aligned with the sheets. The friction is modelled as a shear stress and is depending linearly on the strain rate. That is, $\tau= \mu \partial_y u$ is a friction stress aligned with  the x-direction. Furthermore, in (\ref{visc_BT}) stresses proportional to $u_x$ and $v_y$ appear. These represent  the resistance when the fluid accelerates in the flow direction. (Expansions and contractions.) 

Returning to the control volume, a force that changes $\rho u$ must be directed in the x-direction. At the boundaries $x=0,1$, $\mu u_x$ is such a force. Hence, it is not unreasonable to include it in the model. 

\begin{figure}[ht]
\includegraphics[width=5cm]{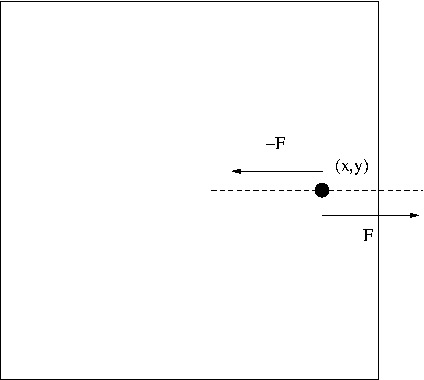}
\caption{A 2-D control volume (in the $xy$-plane) with viscous friction along dashed line.}
\label{fig00}
\end{figure}

To cover all possibilities, we also consider the effect of $\mu u_y$ on an $x$-boundary. As already mentioned, this force is directed in the x-direction. At any point in the fluid, one sheet will exert a force on the neighbouring fluid sheet. However, there will be an equally large force, but with opposite sign, on the other sheet. (See Fig \ref{fig00}.)
The two forces cancel at every point on the sheet and this is also true at the control volume boundary. Hence, there is no contribution to the change of $x$-momentum from $\mu u_y$ at the $x$-boundary.  (Note that this term is not part of the boundary terms in the $x$-direction.) However, on a $y$-boundary it is clear that the sheets just outside and inside the domain create a net force. Therefore, it is not unreasonable that $\mu u_y$ appears on a $y$-boundary, as it does in (\ref{visc_BT}). 

Next, we consider $v$ and its derivatives. In (\ref{visc_BT}), $\mu v_y$ appears on the boundaries, $x=\{0,1\}$. However, $\mu v_y$ is a force tangential to the boundary and it can not affect $\rho u$. 

Furthermore, $\mu v_x$ is a shear force between sheets of fluid aligned with the y-axis. Similarly to $\mu u_y$, the friction force exerted by the sheets on each other cancel in the interior of the domain. However, the force is perpendicular to the x-axis and can not change the $\rho u$ momentum, neither on $x$- nor on $y$-boundaries.

We arrive at the following conclusion: Based on the standard interpretation of the Newtonian model, it is impossible to motivate the full stress tensor in an Eulerian framework. The only terms that could affect $\rho u$ are:
\begin{align}
\mu u_x,\quad x=\{0,1\},\quad
\mu u_y,\quad y=\{0,1\}.\nonumber
\end{align}

\subsection{Comparison with continuum mechanics}

The stress tensor for a perfect Newtonian gas is mathematically similar to that in solid mechanics, but the physics is quite different. In a gas,  diffusion, as in random movement of molecules, causes the forces that are modelled by the stress tensor.

In solid mechanics, the stress tensor is structurally similar for isotropic materials but
the stresses are caused by electromagnetic forces, not diffusion. Hence, a mass parcel is naturally defined without the need for any mass to cross its boundary to generate the forces. In this case, a balance of forces is the fundamental principle and the Lagrangian frame is the natural perspective.

\subsection{Some preliminary conclusions}

In Section \ref{sec:critical}, a number of unphysical properties of the Navier-Stokes equations were discussed. These were  consequences of the velocity definition that  models diffusive movements as well as advective. (This appears to be the same observation as Brenner made, \cite{Brenner05_2}, and just as we do here, he suggested that diffusive mass movements should be separated from the velocity.) The reason that causes the Navier-Stokes equations to mix advective and diffusive transport is  \emph{ the Lagrangian derivation and the balance of forces which is \emph{not} a first principle.} Our observations so far has lead us to a main conclusion in this article.

\begin{postulate}\label{postulate_eulerian}
The inviscid Euler equations should be regularized by a diffusion model (first cause).
\end{postulate}
The natural frame to model diffusion is the Eulerian, and a new model for compressible flows is derived in Section \ref{sec:new_model}. However, some properties of the macroscopic variables in an Eulerian frame must first be established. 

\section{Macroscopic field variables}\label{sec:av_var}

At a microscopic level, a compressible ideal fluid is modelled as a collection of perfectly elastic particles governed by Newtonian mechanics. Herein, we assume that no outside forces affect the system such that the particle system has the following properties:
\begin{itemize}
\item It conserves mass, momentum,  kinetic energy and angular momentum. 
\item Its centre of mass moves uniformly. 
\item It is Galilean invariant.
\item Conservation of angular momentum implies rotational symmetry. 
\end{itemize}
We immediately conclude that a macroscopic continuum model must be Galilean invariant and rotationally symmetric or else the physics would differ between inertial frames. 

In both the macroscopic frames, the Eulerian control volume and the Lagrangian fluid parcel, it is assumed that the macroscopic field variables are known. However, the field variables are obtained as averages over much smaller volumes. These volumes may also be Lagrangian or Eulerian and the distinction may be important. 

\subsection{Density}

\begin{figure}[h]
\includegraphics[width=5cm]{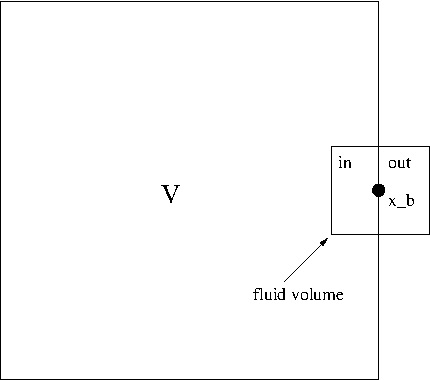}
\caption{A 2-D control volume (in the $xy$-plane) with boundary point and its averaging neighbourhood.}
\label{figV}
\end{figure}

Since our aim is to satisfy Postulate \ref{postulate_eulerian}, we adopt an Eulerian view. Hence, we consider a fixed control volume ($V$), as in Fig. \ref{figV}.  
The total mass inside the volume is equated with the in- and outflux of mass through the boundaries, as in Eqn. (\ref{CV_rho}). The macroscopic density $\rho(\xb,t)$ is in turn computed as the average of the mass in a much smaller fixed neighbourhood of $\xb,t$, denoted \emph{fluid volume}. In Fig. \ref{figV} the fluid volume around a boundary point, $\xb_b$,  is depicted. 

\begin{figure}[h]
\centering
\includegraphics[width=6cm]{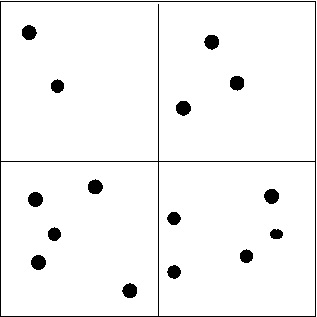}
\caption{Four of the fluid volumes (in the $xy$-plane) that cover the 2-D domain. The black dots symbolizes particles. }
\label{fig_eulerian}
\end{figure}

To define the field variables, we divide space into small squares (or cubes in 3-D), i.e., fluid volumes, see Fig \ref{fig_eulerian}. (We consider 2-D for simplicity and only depict four volumes. These four volumes do not cover the entire space.) 
A fluid volume (subsquare) contains a number of particles and the sum of their  masses divided by the measure of the fluid volume, results in a value for $\rho$  for the fluid volume.

The collection of fluid volumes conserves mass since all particles are accounted for (thanks to the covering of the domain by the fluid volumes) and since a particle belongs to one, and only one, fluid volume. Hence,  the sum of $\rho$-values obtained from the set of fluid volumes is constant.

Note that the fluid volumes are not infinitesimally small since otherwise they would not contain sufficiently many particles to allow a smoothly varying average. Hence, we only obtain values for $\rho$ at discrete points in space meaning that $\rho$ is not a continuum. 
\emph{Therefore, we assume that the discrete function in space is (conservatively) extended to a field in $\Rbb^3$, which allows us to model its evolution by a PDE.} Hence,  each point in (the continuum) space represents a fluid volume and its associated macroscopic field variables can only have the properties that a finite fluid volume has.

\subsection{Momentum}

The same arguments used for mass apply to momentum in a fluid volume. At any instant in time, all particles, and their associated momentum, belong to one and only one fluid volume.  Hence, the macroscopic momentum is locally (and globally) conserved.

So far, we have associated with each Eulerian fluid volume a density and momentum and these quantities are conserved. To continue the discussion, we must specify where these values are located. There are two options: either we interpret $\rho$ and $\rho \vv$ as mean values for the entire fluid volume or we  associate them with an arbitrarily chosen point inside the volume. The two views are equivalent since neither contain more information than the other. 

\begin{figure}[h]
\centering
\includegraphics[width=6cm]{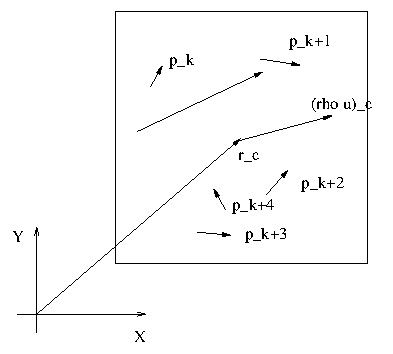}
\caption{A fluid volume and its particles.}
\label{fig_ang_mom1}
\end{figure}

In our discussion, it is more convenient to associate the macroscopic variable with the midpoint of the fluid volume. See Fig. \ref{fig_ang_mom1}, where the square represents one fluid volume. $r_c$ is the vector to the centre point, which we define to be the location of $\rho(r_c,t)$ and $(\rho \vv)(r_c,t)=(\rho \vv)_c$. Furthermore, each arrow represents a particle with a momentum $\pb_k$. Hence, with $N$ particles inside the fluid volume, we have
\begin{align}
(rhou)_c=(\rho \vv)_c= \sum_{i=1}^N \pb_i.\nonumber
\end{align}

\subsection{Energy}

Turning to the particle kinetic energy, it is clear from   Fig. \ref{fig_eulerian} that it is conserved since a particle carries its kinetic energy across  a fluid volume boundary. However, and unlike mass and momentum, the particle kinetic energy in a fluid volume is, in general, not equal to the macroscopic kinetic energy since
\begin{align}
\frac{|\rho \vv|^2}{2\rho}\neq  \sum_{i=1}^N \frac{|\pb_i|^2}{2m}.\label{kin_en}
\end{align}
Hence, we can not infer conservation of the macroscopic kinetic energy from conservation at the particle level. 

The lack of macroscopic conservation is a consequence of $\rho \vv$ not encoding all the information of the particle system. By introducing another macroscopic variable, internal energy (and hence temperature), it is possible to recover macroscopic conservation of the total energy, i.e., the sum of internal and macroscopic kinetic energy.

\subsection{Angular momentum}\label{sec:ang_mom_particles}

\begin{figure}[h]
\centering
\includegraphics[width=6cm]{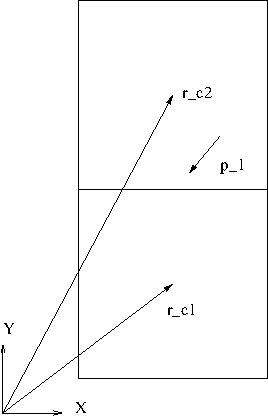}
\caption{Two fluid volumes with particle passing in across boundary.}
\label{fig_ang_mom2}
\end{figure}

Consider an Eulerian fluid volume with  $N$ particles. (See Fig. \ref{fig_ang_mom1}.) All particles have the same mass, $m_p$, and the $i$th particle has momentum $\pb_i$ and position vector $\rr_i$. (See Fig.  \ref{fig_ang_mom1}.) As long as no particle escapes the volume, the total angular momentum is conserved. That is,
\begin{align}
\sum_{i=1}^N \rr_i\times \pb_i=constant.\nonumber
\end{align}
Moreover, if the particle $\pb_1$ in Fig. \ref{fig_ang_mom2} moves from the upper to the lower volume, the decrease of angular momentum in the upper is exactly the same as the increase of the lower. At the particle level, angular momentum is conserved (which is well-known). 

 The macroscopic angular momentum  is  $\rr_c\times (\rho \vv)_c$, where $\rr_c$ is the vector to the centre-point of the volume (see Fig. \ref{fig_ang_mom1}) and $(\rho\vv)_c=\sum_{i=1}^N{\bf p}_i$. In general, we have
\begin{align}
\rr_c\times (\rho \vv)_c\neq \sum_{i=1}^N \rr_i\times \pb_i. \label{noncons_am}
\end{align}

Furthermore, when a particle crosses an interface between two control volumes the change of macroscopic angular momentum in the two volumes does not cancel. To see this, consider Fig. \ref{fig_ang_mom2}. The contribution to the angular momentum of the particle $\pb_1$ to the upper volume is $\rr_{c2}\times \pb_1$. The moment it crosses the boundary to the lower volume, the angular momentum of the upper volume decreases by this amount and the lower increases by $\rr_{c1}\times \pb_1$. The two are not equal. (The same argument is applicable if many particles pass at the same time. However, then it is the average momentum of the particles that plays the role of $\pb_1$. It does not help the matter.) \emph{We conclude that macroscopic angular momentum is not conserved.} 

In (\ref{noncons_am}), we see that in order to model conservation of angular momentum, we need another macroscopic vector-variable which represents the random component of angular momentum within a fluid volume. That is, an ``internal angular momentum density''. This is completely analogous to the role that temperature plays in conservation of energy. The three new components require that the system is extended by three new equations. Such an extended model would have the capability to predict the evolution of mass, momentum, energy and angular momentum. 

 In view of the above arguments, it is \emph{unphysical} to demand that the angular momentum based on the averaged momentum is conserved in an Eulerian compressible continuum model (for mass, momentum and energy), since such a property does not follow from the particle system. (In the inviscid case, the mean flow happens conserve angular momentum since no random motions are assumed to exist in the flow such that  the averages carry all information.)

The only thing the non-conservation of angular momentum means is that this information can not be drawn from the solution. It does not imply that the solution is an inaccurate description of the other three quantities.

\subsection{Centre-of-mass movement}

By similar arguments, it is also evident that the macroscopic continuum variables do not carry enough information to ensure a uniform centre-of-mass movement since the centre-of-mass for each fluid volume is not known due to random movements within the fluid volume.

As all the previous arguments, this is a profound property not connected to any equations modelling the flow. To see this, assume that the positions and velocities are known for all the particles in a system such that the $\rho$-averages of the fluid volumes of Fig. \ref{fig_eulerian} can be computed. Then, the centre-of-mass of the resulting field, $\rho$, does not move uniformly although the centre-of-mass of the swarm of particles does. (The deviation from a uniform movement represents the uncertainty about the state beyond the continuum scale.)

\section{A new Eulerian model}\label{sec:new_model}


In the Navier-Stokes equations, the random motions were responsible for the viscosity whose effect was modelled in the stress tensor and heat conduction. However, in Section \ref{sec:stress}, it was shown that the full viscous stress tensor is not consistent with a control volume view (since all terms can not represent forces). Below, we shall see that the stress tensor violate conservation on the sub-continuum scale. (Therefore, we discard the stress tensor entirely.) However, the heat conductive term of (\ref{NS}) was derived in a control volume framework. Nevertheless, we will see that it has no place in a new model derived in an Eulerian frame. Instead, we arrive at a new model at the end of Subsection  \ref{sec:CV_derivation}.

\subsection{Control volume derivation}\label{sec:CV_derivation}

With the field variables defined (as in the previous section), we define an Eulerian control volume with the aim to calculate the flow in and out of it based on the macroscopic variables. By definition the macroscopic variables do not encode random motions and we obtain the Euler equations,
\begin{align}
\partial_t \rho + div_\xb(\rho \vv )&= 0, \nonumber \\
\partial_t (\rho \vv) + div_\xb(\rho \vv \otimes \vv) + \nabla_\xb p&=0, \label{euler} \\
\partial_t (E) + div_\xb(E \vv +p\vv)  &=0.\nonumber
\end{align}

To model random motions not resolved by the averaged variables, we have to study sub-fluid volume effects. In Fig. \ref{figV}, we denote the macroscopic vector variable on the right boundary point as $\uu(\xb_b,t)$. Next, we split the fluid volume (surrounding $\xb_b$), in two equal halves along the control volume boundary. We average the outer and inner halves separately leading to macroscopic values: $\rho_{out}$ and $\rho_{in}$ (and similarly for the other variables). In general $\rho_{out,in}\neq \rho(\xb_b,t)$ and
\begin{align}
\frac{1}{2}(\rho_{out}+\rho_{in})= \rho(\xb_b,t).\nonumber
\end{align}
 If $\rho_{out}\neq \rho_{in}$, the difference disappears when the gas settles to thermodynamic equilibrium locally. The mechanism for this is diffusion and we assume that it is proportional to $\rho_{out}-\rho_{in}\approx \frac{\delta x}{2}\partial_x \rho$. Furthermore, the speed at which this takes place may depend on the temperature (the random speed of the particles) and density. Hence, on the macroscopic level it is reasonable to model the diffusion within the fluid volume by
\begin{align}
\nu(\rho,T)\partial_n \rho.\nonumber
\end{align}
\begin{remark}
Note that we do not suggest that the macroscopic velocity is responsible for redistributing mass within the fluid volume.
\end{remark}
\begin{remark}
In hindsight, we should have defined the sub-fluid volumes to be the actual fluid volumes and defined the macroscopic variables to be averages of a few fluid volumes. Our definition is valid as long as a sub-fluid volume is large enough to allow smoothly varying averages.
\end{remark}
Turning to momentum, we first consider the Navier-Stokes equations for which the diffusive contributions to the momentum balance are $\mu$ times velocity derivatives. In Section \ref{sec:eulerian_ns}, we argued that most of the terms of the stress tensor can not affect their corresponding momentum components. However, the Laplacian part of the stress tensor could have an affect. Hence, consider the case when $\vv_{in}=\vv_{out}$. Then there is no diffusion, despite that in general $\mm_{in}\neq \mm_{out}$ since mass need not be equally distributed. According to the Navier-Stokes equations (\ref{NS}), a higher number of molecules on one side of the fluid volume does not lead to diffusion of momentum. (Recall that within a fluid volume diffusion should dominate and drive the particles to a local thermodynamic equilibrium.)

Next, let us consider the isolated effect of diffusing velocity. To this end, consider a fluid volume split in two pieces as in Fig. \ref{figV}. On the left there are two particles (of mass $m_p$). One has velocity $2u_p$ and the other $u_p$. Their mean velocity is $1.5u_p$ and their combined momentum is $3m_pu_p$. On the right we have one particle at rest. (Mean velocity and momentum are zero). Hence, the total momentum of all three particles is $3m_pu_p$. Diffusing the mean velocity implies that all particles have velocity $0.75u_p$ and the total momentum is $2.25m_pu_p$. That is, momentum is not conserved at sub-continuum scale by this diffusive mechanism.

\begin{remark}
Momentum would be conserved by velocity diffusion if both halves contain the same mass. That is, if $\rho$ is uniform. This is satisfied for an incompressible fluid where momentum diffusion and velocity diffusion are equivalent.
\end{remark}

 Of course, the Navier-Stokes equations do conserve momentum. The equations counter the above process via the macroscopic velocity that facilitates the transport of mass at all scales. In this way, the Navier-Stokes guarantees mass and momentum conservation.  However, this mechanism is arguably a complicated way to model conservation; it introduces the ambiguous definition of velocity; and it is unphysical since diffusion should dominate on sub-continuum scales (where the advective velocity is not even defined). 

Hence, we discard the stress tensor entirely. We claim that a more physical model is that momentum is directly diffused. Clearly, as particles move randomly between the sub-fluid volumes they bring with them their, mass, momentum and energy, which are then automatically conserved. This is a much simpler description of the random movements.  This argument also implies that the diffusive coefficient for momentum (and later energy) is the same as for mass. That is, we model momentum diffusion by
\begin{align}
\nu(\rho,T)\partial_n \mm.\nonumber
\end{align}

\begin{remark}
We emphasize that viscous stresses could still be proportional to $\mu \nabla_\xb \vv$, when measured in an experiment. We are only claiming that these terms are not the correct terms in the equations for a compressible fluid in an Eulerian frame.
\end{remark}

Having noticed that it is the conserved entities in the particle system that model diffusion at the macroscopic level, we can address the energy equation. The kinetic energy of the particle system is encoded as the total energy at the macroscopic level. (Sum of internal energy and macroscopic kinetic energy.) Without repeating the arguments, it is the gradient of the total energy that drives the energy diffusion and ensures sub-fluid volume conservation. That is,
 \begin{align}
\nu(\rho,T)\partial_n E. \nonumber
\end{align}
\begin{remark}\label{rem_fourier}
Note that temperature diffusion by Fourier's law, is non-conservative with respect to energy in a fluid volume in the same way as the velocity is with respect to momentum. Adding Fourier's law to a model induces micro-advection to enforce conservation. (See also below.)
\end{remark}

If we accept that at every boundary, the normal derivative of the conservative variable is the only (diffusive) flux that changes the variable inside the control volume, we obtain, with the application of Gauss' theorem, the following system,
\begin{align}
\partial_t \rho + div_\xb(\rho \vv )&= \nabla_\xb \cdot (\nu \nabla_\xb \rho),\nonumber \\
\partial_t (\rho \vv) + div_\xb(\rho \vv \otimes \vv) + \nabla_\xb p&=\nabla_\xb \cdot
(\nu \nabla_\xb \rho \vv)), \label{eulerian} \\
\partial_t (E) + div_\xb(E \vv +p\vv)  &= \nabla_\xb \cdot( \nu \nabla_\xb E), \nonumber \\ 
p&=\rho R T, \quad \textrm{ideal gas law,}\nonumber  
\end{align}
where $\nu$ is the diffusion coefficient that may depend on the thermodynamic variables ($\nu=\nu(\rho,T)$) but not on velocity.

We note that (\ref{eulerian}) is consistent with (\ref{new_cont}). As discussed in Section \ref{subsec:pos}, the mass diffusive terms paves the way for a positivity proof. In the following sections, we prove that (\ref{eulerian}) satisfies Postulate \ref{assump1} and has the necessary symmetries. Furthermore, we show that (\ref{eulerian}) has favourable entropy properties and we briefly discuss adiabatic walls and far-field boundaries.

\subsection{Relaxation to steady state.}

In (\ref{eulerian}), heat diffusion by Fourier's law is not included. (Despite that it is derived in an Eulerian view.) Here, we will show that if Fourier's law is added, the  system is not able to relax to thermodynamic equilibrium without inducing a velocity (breaking Postulate \ref{assump1}).

We return to our thought experiment in Section \ref{sec:relaxNS} (and to one space dimension) where (\ref{eulerian}) takes the form:
\begin{align}
\rho_t &=( \nu \rho_x )_x, \nonumber\\
m_t +(\frac{m^2}{\rho}+p)_x &=(\nu m_x)_x , \label{steady_relax} \\
E_t + (\frac{m}{\rho}(E+p))_x  &=(\nu E_x)_x.\nonumber
\end{align}

We  apply the perturbation (\ref{steady_relax}). We demand that $p_t=p_x=0$ as well as $m_t=m_x=0$. At $t=0$, the equations are
\begin{align}
\rho_t &=( \nu \rho_x )_x, \nonumber\\
m_t  &= 0, \label{steady_relax2} \\
\frac{p_t}{\gamma-1}  &=\left(\left( \frac{\nu p_x}{(\gamma-1)}\right)\right))_x,  \nonumber
\end{align}
where we have used that $E=\frac{p}{\gamma-1}$ at the initial
state. In fact, since $p_t=p_x=0$, the last equation vanishes entirely. The relaxation is only facilitated by the mass diffusion in the first equation. Furthermore, since $p$ is constant, the temperature relaxes as a consequence of the gas law.

Clearly, an explicit temperature diffusion, $(\kappa T_x)_x$, in the energy equation, destroys the property of relaxation (as pointed out in Remark \ref{rem_fourier}). It implies that $p_t\neq 0$, which induces a momentum. (Hence, we discard Fourier's law.) However, we stress that this does not mean that there is no temperature diffusion, only
that it is connected to the mass-diffusion term. The right-hand side of the energy equation in (\ref{steady_relax2}) can be rephrased as
\begin{align}
\frac{\nu p_x}{\gamma-1}  &=  \nu c_v(\rho_xRT+\rho R T_x). \nonumber
\end{align}
Hence, the heat flux is
\begin{align}
\nu c_v\rho T_x.  \label{temp_diff}
\end{align}

In the standard Navier-Stokes model, it is only temperature that changes the internal energy by a diffusive process. In a flow where $\rho_x>>1$ and $T_x \approx 0$ there would be no diffusion of internal energy. The strong gradient of $\rho$ would not imply that molecules randomly travel to neighbouring volumes carrying their internal energy with them. This mechanism would be facilitated, as has been mentioned several times, by an induced microscopic flow that advects the internal energy. In the new system the diffusion acts directly on the internal energy.

\subsection{Galilean invariance}\label{sec:galilean}

The system (\ref{eulerian}) can be cast in divergence form:
\begin{align}
\uu_t + \fb(\uu)_x  + \gb(\uu)_y  + \hb(\uu)_z=                                        \Fv^m(\uu)_x  + \Gv^m(\uu)_y  + \Hv^m(\uu)_z\label{con_form}
\end{align}
where $\uu=(\rho, \rho \vv^T, E)^T$ and
 \begin{align}
 \Fv^m=\nu \partial_x \uu,\quad
 \Gv^m=\nu \partial_y \uu,\quad
 \Hv^m=\nu \partial_z \uu.\label{diff_flux} 
 \end{align}
Furthermore, $\fb,\gb,\hb$ denote the inviscid fluxes. To reduce notation, we use two space dimensions. Let $U$ denote a (constant) velocity difference between the two inertial frames $(x,y)$ and $(\xi,\eta)$. The Galilean transformation is:
\begin{align}
x=\xi+Ut,\quad
y=\eta,\quad
t=\tau,\quad
u=u'+U,\quad 
v=v',
\nonumber \\
\partial_x=\partial_\xi,\quad
\partial_y=\partial_\eta,\quad
\partial_t=\partial_\tau-U\partial_\xi.\nonumber 
\end{align}

The continuity equation 
\begin{align}
\rho_t + (\rho u)_x+(\rho v)_y= (\nu\rho_{x})_x+(\nu\rho_{y})_y.\nonumber 
\end{align}
transforms to
\begin{align}
\rho_\tau+ (\rho u')_\xi+(\rho v')_\eta=(\nu\rho_{\xi})_\xi+(\nu\rho_{\eta})_\eta.\label{gal_cont}
\end{align}
Hence, the continuity equation is Galilean invariant.

The x-momentum equation is, 
\begin{align}
(\rho u)_t + (\rho u^2+p)_x+(\rho vu)_y&=(\nu(\rho u)_{x})_x+(\nu(\rho
  u)_{y})_y.\nonumber 
\end{align}
Applying the transformation and using (\ref{gal_cont}) result in
\begin{align}
(\rho u')_\tau  
+ (\rho ((u')^2)+p)_\xi+( \rho v' u'))_\eta=
(\nu(\rho u')_{\xi})_\xi +(\nu(\rho u')_{\eta})_\eta.\nonumber 
\end{align}
In the same way we transform the y-momentum,
\begin{align}
(\rho v)_t + (\rho uv)_x+(\rho v^2+p)_y=(\nu(\rho v)_{x})_x+(\nu(\rho v)_{y})_y,\nonumber 
\end{align}
and obtain
\begin{align}
(\rho v')_\tau + (\rho u'v')_\xi+(\rho (v')^2+p)_\eta=(\nu(\rho  v')_{\xi})_\xi+(\nu(\rho v')_{\eta})_\eta.\nonumber
\end{align}
We conclude that both momentum equations are Galilean invariant.

The energy equation is,
\begin{align}
E_t+ (u(E+p))_x +(v(E+p))_y=(\nu E_{x})_x+(\nu E_{y})_y.\nonumber 
\end{align}
which transforms to,
\begin{align}
E'_{\tau}+(u'(E'+p))_{\xi}
+(v'(E'+p))_\eta
&=
(\nu E'_{\xi})_\xi+(\nu E'_{\eta})_\eta,\nonumber
\end{align}
where $E'=\frac{1}{2}\rho((u')^2+(v')^2)+\frac{p}{\gamma-1}$. (Details of the derivation are found in Appendix \ref{app:galilean}.) We conclude that (\ref{eulerian}) is Galilean invariant.

\subsection{Rotational symmetry}\label{sec:rot_symm}

Since the orientation of the (Cartesian) coordinate system in which we formulate our physical laws is arbitrary, we must demand that the equations take the same form after any rotation. This is well-known to hold for the inviscid Euler equations and we only need to consider this property for the viscous terms of (\ref{eulerian}). 

A rotation by  an angle $\theta$ in the $xy$-plane (around the $z$-axis), is expressed as a change of coordinates:
\begin{align}
\left(\begin{array}{c} \xi \\ \eta \\ \zeta \end{array}\right)=
\left(\begin{array}{ccc} \cos \theta  & -\sin \theta & 0 \\
\sin \theta  & \cos \theta & 0 \\
0 & 0 & 1
 \end{array}\right)
\left(\begin{array}{c} x \\ y \\ z \end{array}\right),\nonumber
\end{align}
which leads to the following relations:
\begin{align}
\partial_x &= \phantom{-}\cos \theta \,\partial_\xi +\sin \theta \,\partial_\eta ,\nonumber\\
\partial_y &= -\sin \theta \,\partial_\xi +\cos \theta\, \partial_\eta ,\nonumber \\
\partial_z&=\phantom{-}\partial_\zeta .\nonumber
\end{align}
It is straightforward to apply the transformation and obtain,
\begin{align}
(\nu \rho_x)_x+(\nu \rho_y)_y+(\nu \rho_y)_y=
(\nu \rho_\xi)_\xi+(\nu \rho_\eta)_\eta+(\nu \rho_\zeta)_\zeta .\nonumber
\end{align}
The transformations of momentum and energy diffusion are the same. Since the choice of rotation around the $z$-axis was arbitrary, the same result holds for rotations around any of the three axes. Observing that a general rotation, around all the three axes, is a consecutive application of the three possible rotations, we conclude that (\ref{eulerian}) is rotationally symmetric.

\subsubsection{Angular momentum and centre-of-mass movement}

Previous attempts to  modify the Navier-Stokes systems have been rebutted for lack of rotational symmetry. (See \cite{Ottinger09}.) Therefore, we make a few more remarks regarding rotational symmetry. For a system of particles, conservation of angular momentum is equivalent to rotational symmetry. In \cite{Ottinger09}, it was shown that a class of modifications of the Navier-Stokes equations did not preserve angular momentum and hence they concluded that the system is not rotationally symmetric.

In the Eulerian frame, (\ref{eulerian}), the conservation of particle angular momentum does not imply that the macroscopic angular momentum is conserved. Nevertheless, the macroscopic system has to be rotationally symmetric in the sense shown above (and the system (\ref{eulerian}) is.)

As an observation, we remark that when $\nu=constant$ in (\ref{eulerian}), the macroscopic angular momentum \emph{is} conserved. (See Appendix \ref{sec:ang_mom}.) This is because the diffusion becomes perfectly isotropic. However, this property is \emph{not} equivalent to conservation of angular momentum  for the particle system since the internal angular momentum is not accounted for.

In \cite{Ottinger09}, modified Navier-Stokes models were required to allow a  uniform centre-of-mass movement. If $\nu=\nu(\rho)$ is integrable, (\ref{eulerian}) supports a uniform centre-of-mass movement thanks to sufficient symmetry in the diffusion. (See Appendix \ref{sec:Booster}.) However,  $\nu=\nu(\rho,T)$ in general and we have already shown that the uniform centre-of-mass movement in the particle system does not carry over to the macroscopic Eulerian system. Hence, one can not require that the Eulerian system (\ref{eulerian}) has this property.

\subsection{Entropy}

The standard (mathematical) entropy is a convex entropy function $U(\uu)$, and the associated entropy fluxes, $F,G,H$ that satisfy
\begin{align}
\qb^T\fb= F_\uu,\quad \qb^T\gb= G_\uu, \quad \qb^T\hb= H_\uu, \nonumber
\end{align}
where $\qb^T=\frac{\partial U}{\partial \uu}$ is the vector of entropy variables.  Convexity of $U$ implies that $U_{\qb\qb}=\uu_\qb$ is symmetric positive definite.

We multiply (\ref{con_form}) (or (\ref{eulerian})) by $\qb^T$ and integrate in (periodic) space. The result is 
\begin{align}
\int_{\Omega} U_t \,d\xb + \int_{\Omega} (F_x +G_y+H_z)\,d\xb = \int_{\Omega}\qb^T(\Fv_x^m+\Gv_y^m+\Hv_z^m)\,d\xb.\nonumber
\end{align}
By periodicity
\begin{align}
\int_{\Omega} U_t \,d\xb=\int_{\Omega}\qb^T(\Fv_x^m+\Gv_y^m+\Hv_z^m)\,d\xb.\label{ent_est1}
\end{align}
To be entropy consistent the right-hand side must be non-positive. We have
\begin{align}
\int_{\Omega}\qb^T(\Fv^m_x+\Gv^m_y+\Hv^m_z)\,d\xb= 
\int_{\Omega}-\qb^T_x\Fv^m-\qb_y^T\Gv^m-\qb_z^T\Hv^m\,d\xb&=\nonumber \\ 
\int_{\Omega}-\nu\left(\qb^T_x\uu_x+\qb_y^T\uu_y+\qb_z^T\uu_z\right)\,d\xb=
\int_{\Omega}-\nu\left(\qb^T_x\uu_\qb\qb_x+\qb_y^T\uu_\qb\qb_y+\qb_z^T\uu_\qb\qb_z\right)\,d\xb&\leq 0,\nonumber 
\end{align}
since $\uu_\qb$ is positive definite for positive density and temperature.  

\begin{remark}
The Second Law of Thermodynamics requires $\int_{W} S\,d\xb$ to increase on any subdomain $W$ with positive measure, which is equivalent to a decrease of $\int_{W} U\,d\xb$. It is straightforward to first derive an entropy equation and then integrate it against compactly supported test functions, to show that the entropy inequality is also satisfied locally (on any W).  
\end{remark}

Note that we have not assumed a specific $U$. In fact, it follows that the system is entropy diffusive with respect to \emph{all} Harten's generalized entropies, \cite{Harten83}. Hence, (\ref{eulerian}) satisfies the entropy minimum principle, \cite{Tadmor86}. (Mathematically, multiple entropy functions is a useful tool, e.g., when proving boundedness and uniqueness of solutions. The standard system (\ref{NS})  has  neither of these properties. It is only entropy diffusive with respect to the standard entropy,  $U=-\rho S$.) 

For the system (\ref{eulerian}), the precise entropy diffusion for $U=-\rho S$ is
\begin{align}
\qb_x^T\Fv^m=\qb_x(\nu \uu)_x &=\nonumber \\
\left(\frac{1}{c_vT}\left(c_vT(S-\gamma)+\frac{u^2+v^2+w^2}{2}, -u,-v,-w, 1\right)\right)_x(\nu \uu)_x&=\nonumber\\
\nu(\gamma-1)\rho\left( \frac{\rho_x^2}{\rho^2}\right)+\nu\rho\left(
\frac{T_x^2}{T^2}\right)+\nu\frac{\rho}{c_vT}(u_x^2+v_x^2+w_x^2), \nonumber
\end{align}
and similar contributions for the other fluxes. Note that mass diffusion contributes to the entropy diffusion which is not the case for (\ref{NS}).

\subsection{Adiabatic Wall} \label{sec:wall_eul}

The system (\ref{eulerian}) is parabolic and therefore requires five boundary conditions at a wall (and any other type of boundary in three space dimensions). The extra boundary condition must involve density and the natural choice is
\begin{align}
\partial_{\bf n}\rho=0,\nonumber
\end{align}
i.e., there is no diffusive mass flux through the wall.

Carrying out the derivation leading (\ref{ent_est1}) but on a domain $\Omega$ with wall boundaries, results in
\begin{align}
\int_{\Omega} U_t \,d\xb=\int_{\partial \Omega} (F,G,H)\cdot {\nn} \,d\partial \Omega = \int_{\Omega}\qb^T((\Fv^m,\Gv^m,\Hv^m)\cdot {\nn})\,d\partial \Omega\nonumber \\
-
\int_{\Omega}\qb^T_x\Fv^m+\qb^T_y\Gv^m+\qb^T_z\Hv^m\,d\xb.\nonumber
\end{align}
It is straightforward to check  that the inviscid boundary terms cancel by $\vv\cdot \nn=0$ and the diffusive terms by $\vv=0, \partial_nT=0$ and $\partial_n \rho=0$. Hence, $U$ is bounded from above. Furthermore,
\begin{align}
\frac{\partial S}{\partial \nn}=(1-\gamma)\frac{\partial \rho}{\partial \nn}+\frac{\partial T}{\partial \nn}=0,\nonumber
\end{align}
at an adiabatic wall. This is consistent with (\ref{eulerian}) modelling diffusive effects.

As already mentioned, $\partial_n \rho=0$ rules out mass diffusion through the wall and in view of the discussion in Section \ref{sec:adiabatic1} it also ensures that there is no energy diffusion through the wall.

\subsection{Far-field}\label{sec:farf_eul}

In analogy with the Navier-Stokes equations, we linearize and symmetrize (\ref{eulerian}) and obtain a system of the form (\ref{linear_symm}). A major difference from the Navier-Stokes equations is that $B_{ij}$ are non-zero on the first row. 
Hence, the boundary condition
\begin{align}
A^+\ww-\Fv_V&=A^+\gb(t), \quad x=0.\nonumber\\ 
\Fv_V&=B_{11}\ww_x+B_{12}\ww_y+B_{13}\ww_z,\nonumber
\end{align}
is a valid choice irrespective of the flow conditions. Furthermore, by sending $\nu$ to 0, this boundary condition reverts to the Euler conditions (\ref{bc_euler}). 

\subsection{Lagrangian form}

In order to interpret the system (\ref{eulerian}) in a Lagrangian framework, the material derivative has to be adjusted. The velocity is no longer transporting a parcel of constant mass. We rewrite the continuity equation (in 2D to reduce notation),
\begin{align}
\rho_t + (\rho\left(u - \nu (\ln \rho)_x \right))_x+ (\rho\left(v - \nu (\ln \rho)_y \right))_y=0.\nonumber
\end{align}
The effective velocities that transport a mass  parcel are 
\begin{align}
 u - \nu (\ln \rho)_x \quad \textrm{and} \quad 
 v - \nu (\ln \rho)_y.\nonumber
\end{align}
Rewriting the momentum equations of (\ref{eulerian}) to an equation corresponding to (\ref{ns_momentum}), results in
\begin{align}
\rho \left( u_t +(u - \nu (\ln \rho)_x)u_x + (v - \nu (\ln \rho)_y )u_y\right)=-p_x + (\nu \rho u_x)_x+(\nu \rho u_y)_y\label{lagrange_euler}
\end{align} 
On the left-hand side is $\rho$ times the (modified) material derivative of $u$. On the right, is the pressure gradient and the viscous stresses.

\subsection{The diffusion coefficient $\nu$}

In (\ref{lagrange_euler}), we are presented with an opportunity to determine $\nu$. Recognizing that the Navier-Stokes system has been extensively tuned to match experimental data, it is reasonable to connect $\nu$ and $\mu$ such that the diffusive effect in (\ref{eulerian}) is comparable to (\ref{NS}). In (\ref{lagrange_euler}), the term $\nu \rho u_x$ resembles $\mu u_x$ which leads to a diffusion coefficient, $\nu=\frac{\mu}{\rho(x,t)}.$ That is, $\nu$ is the kinematic viscosity. (It may be temperature dependent since $\mu=\mu(T)$.) Furthermore, this choice implies that (\ref{eulerian}) reduces to the standard \emph{incompressible} Navier-Stokes equations when the density is constant.

However, in the Navier-Stokes equations the term $\mu u_x$ is scaled by a factor $4/3$, which may suggest that $\nu=\frac{4\mu}{3\rho(x,t)}$.  Hence, we propose that
\begin{align}
\nu=\alpha\frac{\mu}{\rho(x,t)}+\beta(\rho,T),\label{nu_choice}
\end{align} 
where $\alpha\in [1,4/3]$ according to our deliberations above, but we acknowledge that $\alpha$ could be further adjusted to match experiments. However, we anticipate that the primary dependency is $\nu\sim \mu(T)/\rho$ such that $\alpha$  is a constant. Furthermore, we include another diffusion coefficient $\beta(\rho,T)$ to allow for secondary dependencies.  (In the remaining of this article, we take $\beta=0$.)

\subsection{The Boltzmann equation}

It is well-known that the Navier-Stokes equations can be derived from the Chapman-Enskog expansion of the Boltzmann equation, see \cite{Chapman}. At the time of writing, it is not clear if the model (\ref{eulerian}) can be connected to the Boltzmann equation. One obstacle is that the Boltzmann equation also takes its starting point in a Lagrangian mass element. This leads naturally to the definition of velocity used in the Navier-Stokes equations.

Efforts to introduce mass diffusion in continuum models derived from the Boltzmann equations have been made in \cite{DadzieReese08,DadzieReese10,DadzieReese12,DadzieReese12_2}. Although resembling (\ref{eulerian}) in many ways these models are not exactly the same.  We also mention the approach taken in \cite{Goddard10} that does not its starting point in the Boltzmann equation but derives a continuum model from statistical mechanics in an Eulerian frame.

\section{Numerical experiments}\label{sec:numerics}

Numerical solutions of the two systems, (\ref{NS}) and (\ref{eulerian}), have been computed for three numerical experiments: a steady-state shock, a Kelvin-Helmholtz instability and a blast wave confined between two walls.

\subsection{Steady-state shock.}\label{shock_dynamics}

The initial state is a Mach 2 inviscid steady shock located at $x=0$. The left state is given in SI units by 
\begin{align}
Ma=2,  \quad \rho_L = 1.205, \quad p_L   = 101325,\nonumber 
\end{align}
and the right state is given by the Rankine-Hugoniot relations. The fluid properties are,
\begin{align}
 R =  286.84, \quad  
c_p   = 1005, \quad \gamma   = 1.4, \quad
\mu   = 18.208\cdot 10^{-4},\quad \kappa = 2.57. \nonumber 
\end{align}
For (\ref{eulerian}), we use (\ref{nu_choice}) with $\alpha=1$ and $\alpha=4/3$ (and $\beta=0$).

\begin{figure}[h]
\centering
\subfloat{\includegraphics[width=10cm]{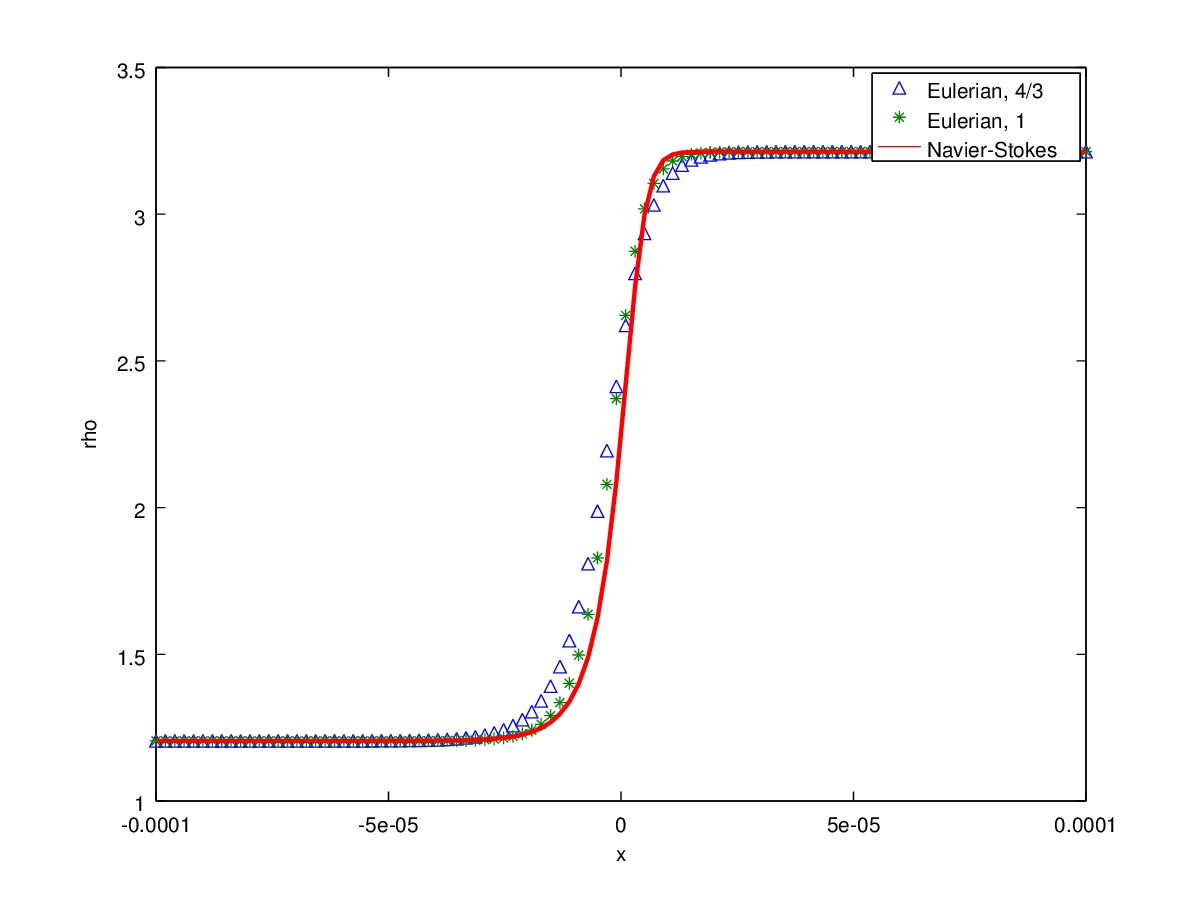}}
\caption{The $\rho$-solution at steady state obtained with 100 grid points.}
\label{fig_rho_steady}
\end{figure}

The schemes used for this problem are described in Appendix \ref{app:scheme}. The density profiles are shown in Fig. \ref{fig_rho_steady}. We conclude that the Eulerian system with $\alpha=1$ is only slightly more diffusive than the Navier-Stokes equations but the difference is not very large. With $\alpha=4/3$, the difference is more pronounced although still not very large.



\begin{figure}[h]
\centering
\subfloat{\includegraphics[width=10cm]{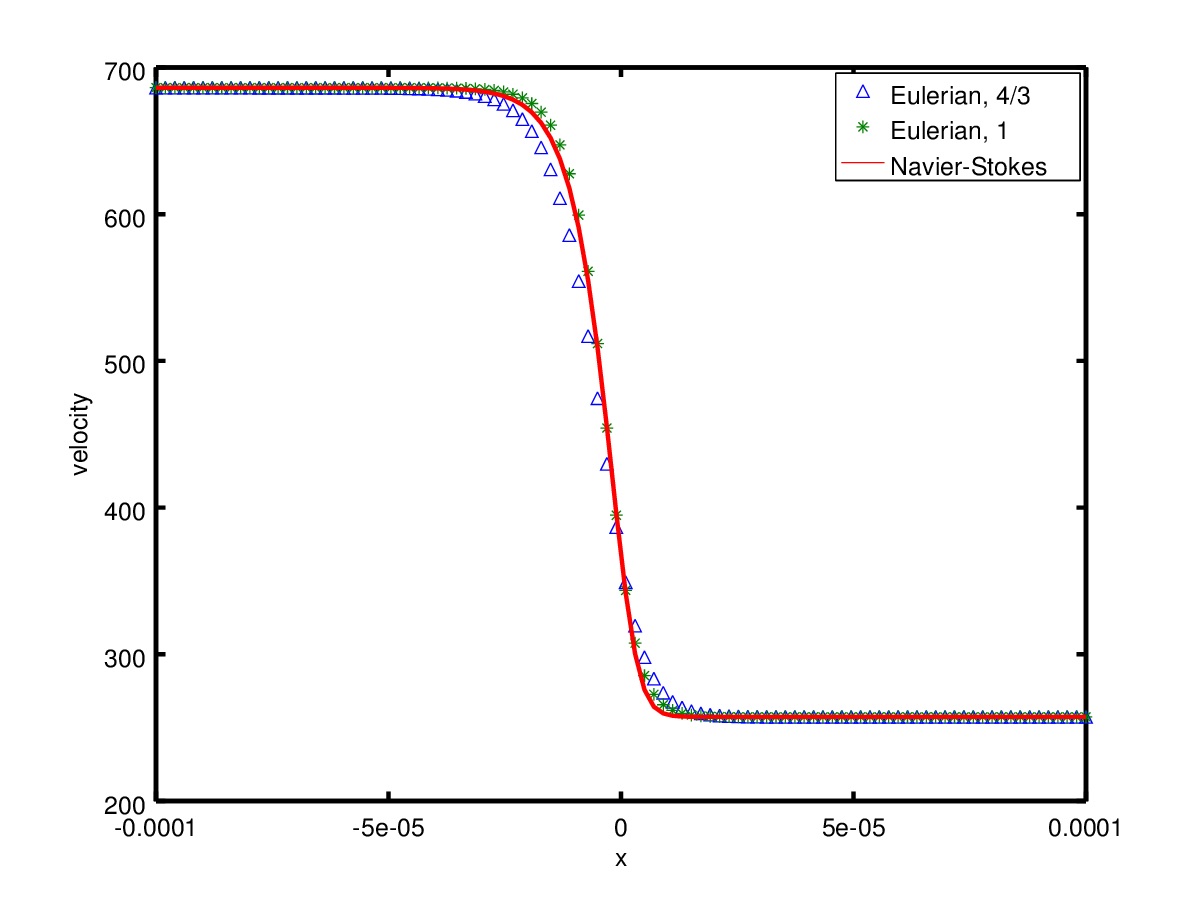}}
\caption{The velocity at steady state obtained with 100 grid points.}
\label{fig_vel_steady}
\end{figure}

The corresponding picture of the velocity is shown in Fig. \ref{fig_vel_steady}. The differences follow the same pattern, but it may be noted that the Eulerian model with $\alpha=1$ has the sharpest velocity profile upstream of the shock. 

However, for the momentum variable the simulations of (\ref{eulerian}) and the Navier-Stokes (\ref{NS}) are significantly different. For the Navier-Stokes equations the momentum is  constant in the entire domain while in the Eulerian simulations it peaks at the shock. The difference between the Eulerian solution and the constant is a measure of the mass diffusive flux. 

\begin{remark}
Note that momentum is not globally conserved in either of the simulations. It is only conserved up to the point where the solution interacts with the boundary, i.e., initial disturbances from the shock impinge on the boundary. At that point the boundary conditions produce an in- or outflux of momentum (and the other conserved variables).  
\end{remark}

\begin{figure}[h]
\centering
\subfloat{\includegraphics[width=10cm]{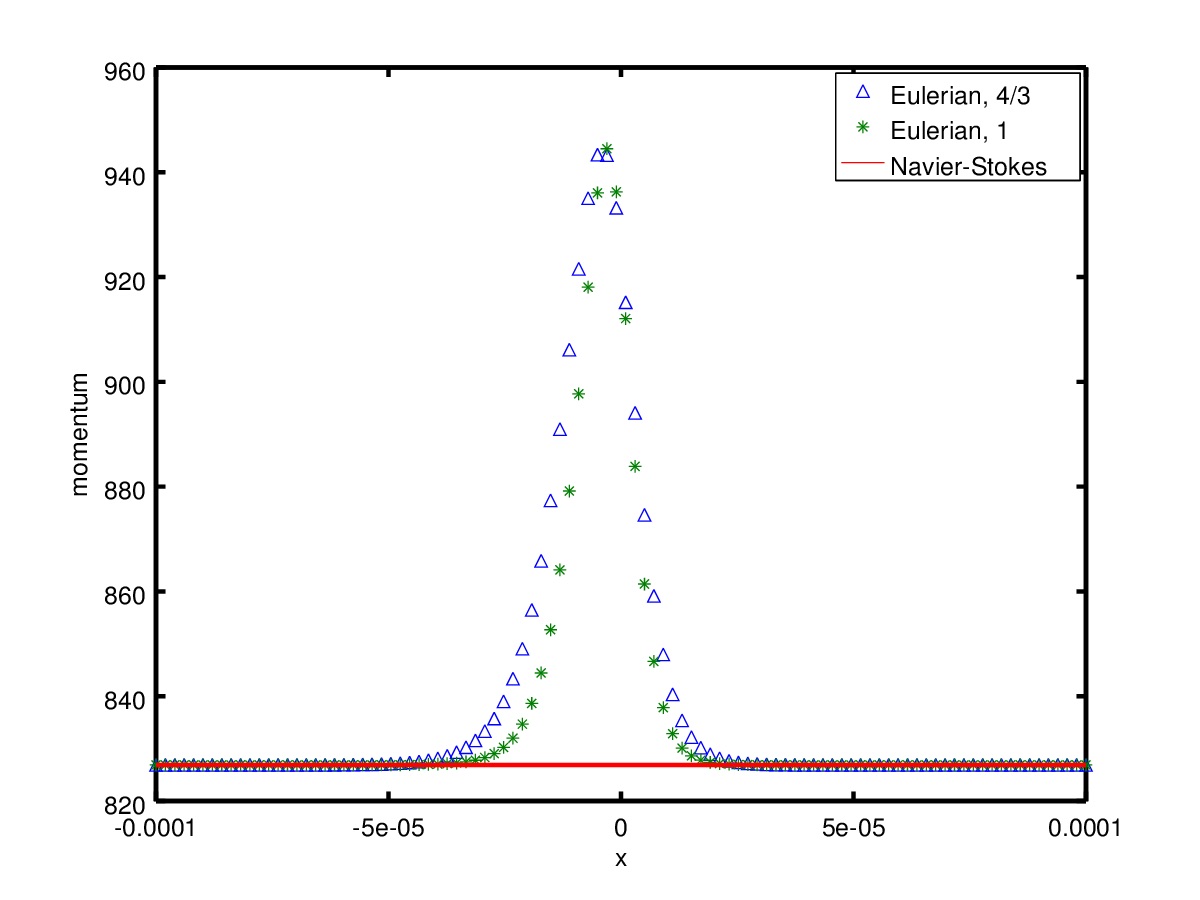}}
\caption{The momentum at steady state obtained with 100 grid points.}
\label{fig_mom_steady}
\end{figure}
\begin{figure}[h]
\centering
\subfloat{\includegraphics[width=10cm]{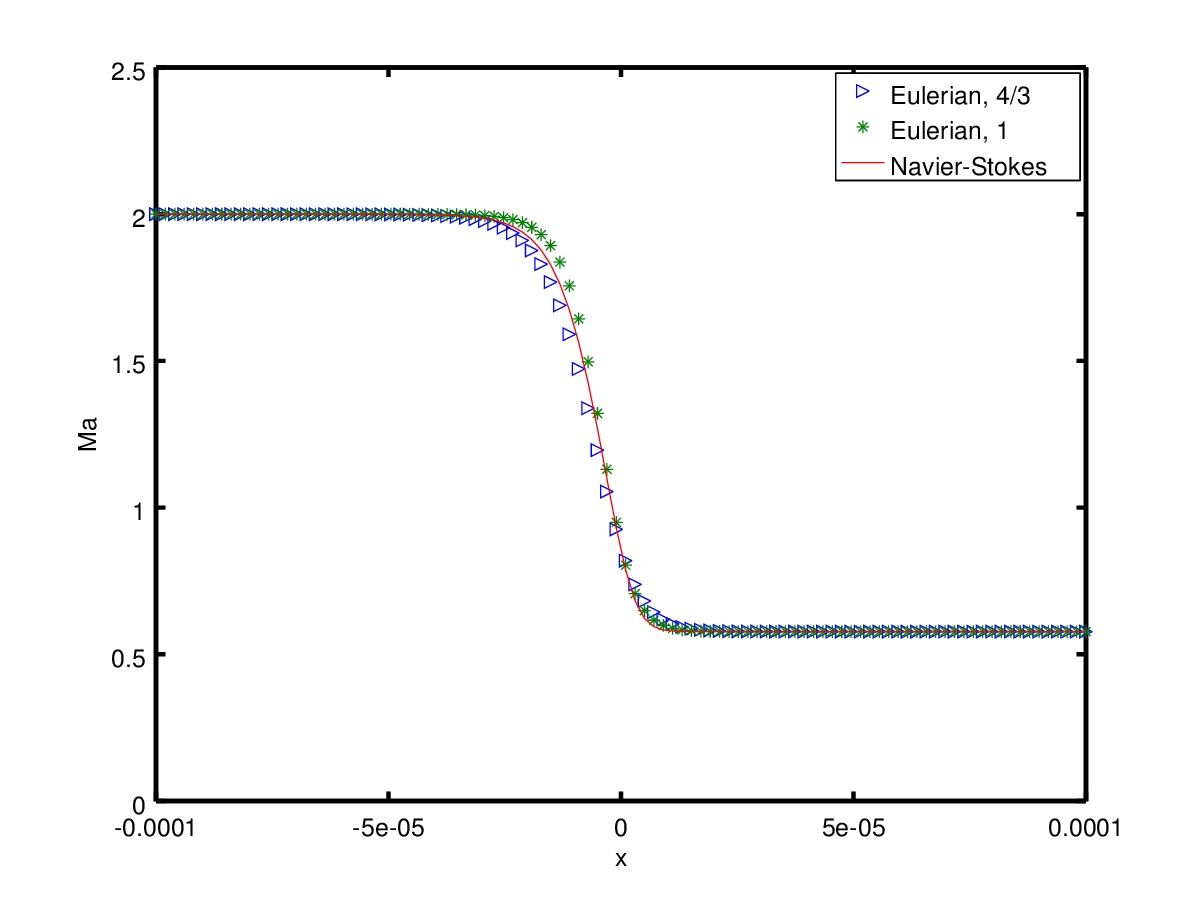}}
\caption{The Mach-number at steady state obtained with 100 grid points.}
\label{fig_Ma_steady}
\end{figure}

The  energy, pressure, temperature all have the same qualitative appearance as the velocity and density. It is only the momentum that differs significantly. However, we also include the figure of the local Mach-number, Fig. \ref{fig_Ma_steady}. We note that neither choice of Eulerian diffusivity results in any overshoot of the Mach-number as was the case for Brenner's modification, \cite{GreenshieldsReese07}. 

\subsection{Kelvin-Helmholtz}\label{sec:KH}

The initial conditions are given by
\begin{equation}
\uu^0=
\begin{cases}
\uu_1 & \text{if }0.25< y<0.75, \\
\uu_2 & \text{if } y\leq 0.25 \text{ or } y\geq 0.75,
\end{cases}\label{initial1}
\end{equation}
where $\uu_1$ and $\uu_2$ are the conservative variables obtained from the
states: $\rho_1=2$ and $\rho_2=1$; $u_1=-0.5+\epsilon \sin(4\pi x)$ and
$u_2=0.5+\epsilon \sin(4\pi x)$; $v_1=v_2=0+\epsilon\sin(4\pi y)$;
$p_1=p_2=2.5$. We use $\epsilon=0.01$ to trip the instability that produce the
familiar Kelvin-Helmholtz roll-ups, and run the simulations till $T=2$.

The fluid properties are
\begin{align}
\mu =2.0\cdot 10^{-4}, \quad 
c_p=1005, \quad 
Pr=0.72, \quad
\kappa = \frac{\mu c_p}{Pr}, \quad 
\gamma  = \frac{c_p}{c_v}=1.4.\nonumber
\end{align}
For the system (\ref{eulerian}), we have used (\ref{nu_choice}) with $\alpha=\{1,4/3\}$ (and  $\beta=0$).

We have run a standard second-order central finite-difference scheme, without any numerical (artificial) diffusion. The only diffusion appearing in the scheme comes from the diffusive terms in the equations. The grid is Cartesian and equidistant with periodic boundary conditions. (See Appendix \ref{app:scheme} for details on the scheme.)

\begin{figure}[ht]
\subfloat[Navier-Stokes]{\includegraphics[width=4.5cm]{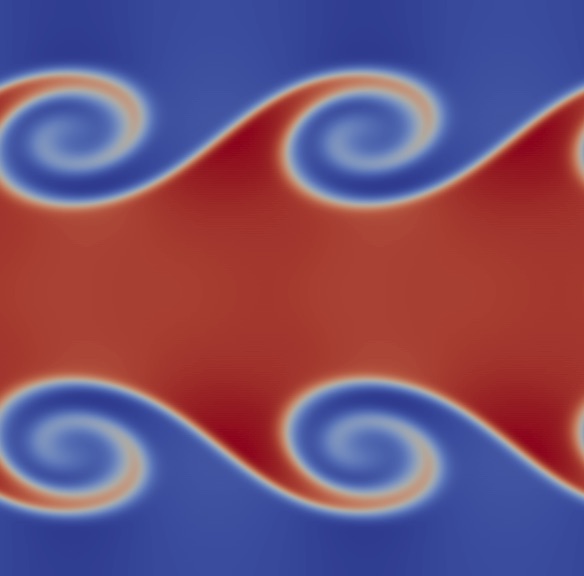}}
\subfloat[Eulerian, $\alpha=1$]{\includegraphics[width=4.5cm]{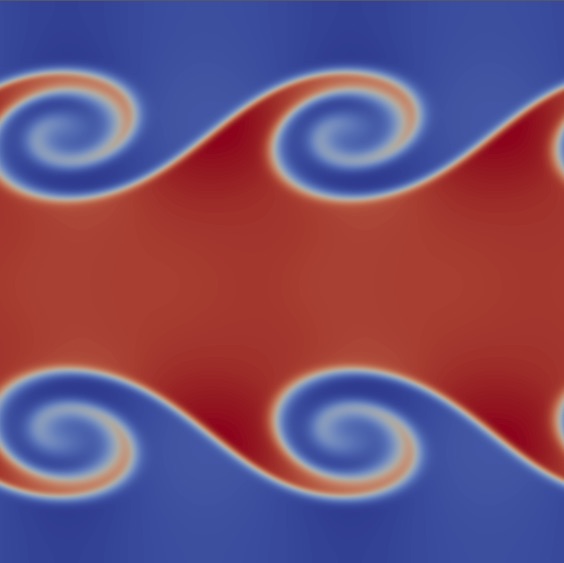}}
\subfloat[Eulerian, $\alpha=4/3$]{\includegraphics[width=4.5cm]{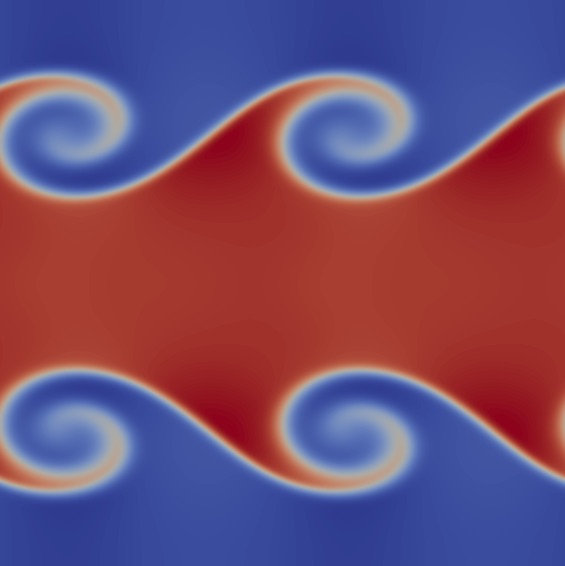}}
\caption{The density at $T=2$ computed with $512^2$ grid points. ($\nu=\alpha\mu/\rho$)}
\label{fig1}
\end{figure}

\begin{figure}[ht]
\subfloat[Navier-Stokes]{\includegraphics[width=4.5cm]{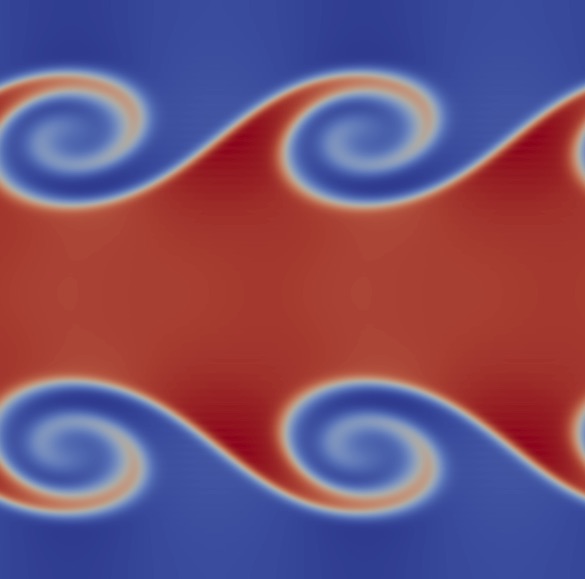}}
\subfloat[Eulerian, $\alpha=1$]{\includegraphics[width=4.5cm]{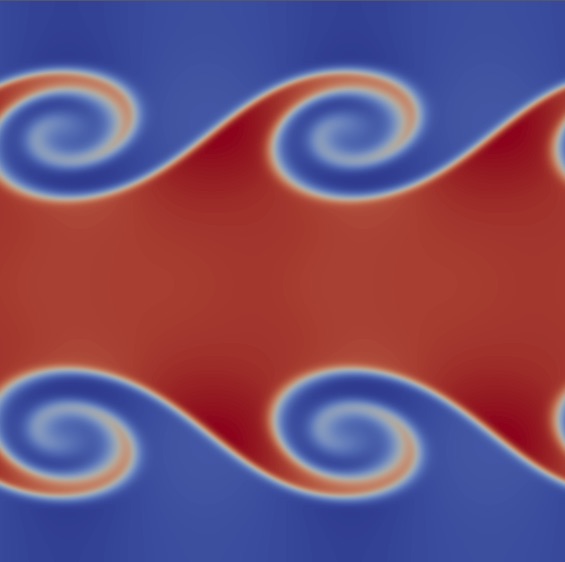}}
\subfloat[Eulerian, $\alpha=4/3$]{\includegraphics[width=4.5cm]{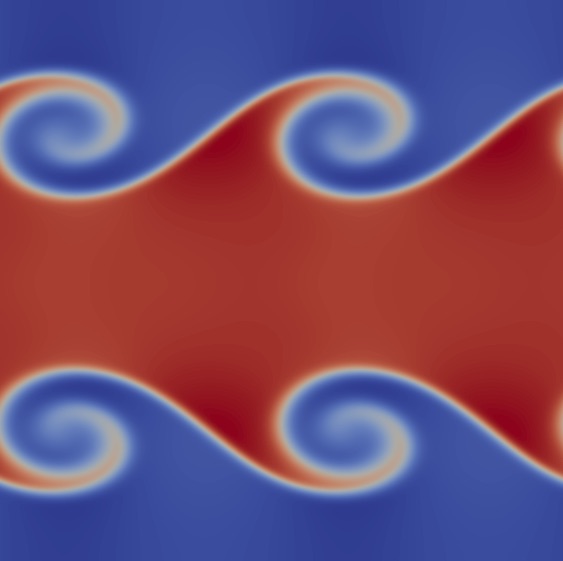}}
\caption{The density at $T=2$ computed with $1024^2$ grid points. ($\nu=\alpha\mu/\rho$)}\label{fig2}
\end{figure}

In Figure \ref{fig1},  the solutions at $T=2$ with $512^2$ grid points are shown. 
The figures are very similar but not exactly the same. This is also the case with $1024^2$ grid points, as depicted in Fig. \ref{fig2}. Judging by the fidelity of the roll-ups, (\ref{eulerian}) with $\alpha=1$ appears to be the least diffusive, the Navier-Stokes equations intermediate and $\alpha=4/3$ the most diffusive. The max and min values of the density are almost the same. On the $1024^2$, density range is $0.96-2.11$ with $\alpha=1$, for the Navier-Stokes $0.97-2.11$, and for $\alpha=4/3$ the range is $0.97-2.10$. The numerical errors might still have an influence, albeit small, on the range. For the other variables the ranges are similarly equal. It appears that the two systems yield strikingly similar results.

\begin{remark}
Other choices of $\nu$ (than $\nu\sim \rho^{-1}$) have a strong impact on the solutions of (\ref{eulerian}) and significantly increase the differences. 
\end{remark}

\subsubsection{ Under-resolved simulations }\label{sec:relax}

Given that the initial data is discontinuous, it is expected that oscillations are produced in the shortest wavelengths on every grid and irrespective of which system that is solved. Since the solutions on fine grids were very similar, it is evident that the diffusive terms in both systems are sufficiently strong to quickly damp these oscillations. However, on a coarse grid the shortest wavelength is longer and the parabolic terms dampen these frequencies less effectively. Hence, the differences between the systems should become more apparent. 

\begin{figure}[ht]
\subfloat[Navier-Stokes]{\includegraphics[width=4.5cm]{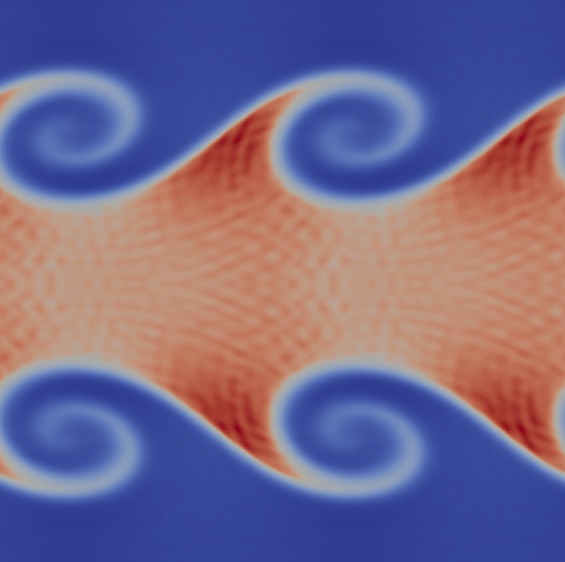}}
\subfloat[Eulerian, $\alpha=1$]{\includegraphics[width=4.5cm]{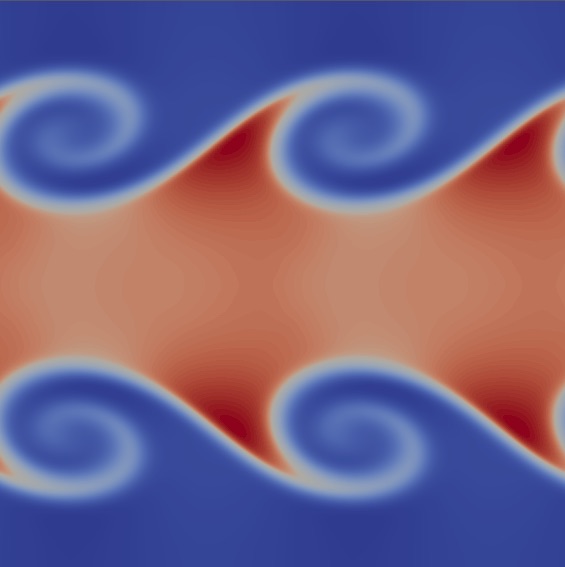}}
\subfloat[Eulerian, $\alpha=4/3$]{\includegraphics[width=4.5cm]{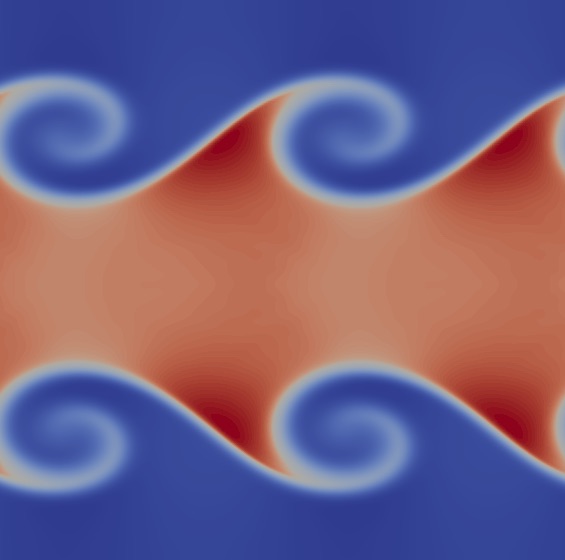}}
\caption{$\rho$ at $T=2$ with $256^2$ grid points. ($\nu=\alpha\mu/\rho$)}
\label{fig3}
\end{figure}

On a  $256^2$ grid the solutions are depicted in Fig. \ref{fig3}.
The Eulerian system (\ref{eulerian}) produces solutions that are very similar to the $512^2$ solution. (The colour mapping is changed to highlight oscillatory structures.) The solutions for $\alpha=1,4/3$ are approximately the same as on the finer grids, which can be seen in the lengths of  the roll-ups. Hence, they are reasonably well resolved.

However, the solution of the standard Navier-Stokes system in Fig. \ref{fig3} is very oscillatory. Although the wave lengths correlate with the grid size, the oscillations are not grid aligned but emanate from the sharp transition between the two fluids. Hence, they are not an artefact of the numerical scheme. The difference between the two models is that (\ref{eulerian}) (for both values of $\alpha$) effectively damps the high-frequency oscillations, although the overall diffusion is not stronger (at least when $\alpha=1$). 

Although it is generally impossible to establish that one phenomenon is a cause and another is an effect in the solution of a non-linear PDE, it is nevertheless likely that the persisting oscillations in the Navier-Stokes solution are connected to the lack of mass diffusion. The fact that the parabolic terms of the Navier-Stokes equations have a zero eigenvalue implies that there is a mode that is not damped. As we have argued, this mode is the redistribution of mass via micro advection. In accordance with Postulate \ref{assump1}, the system (\ref{eulerian}) is designed to allow thermodynamic relaxation which reduces the oscillatory redistribution of mass. This would explain the different behaviour between the two systems. 
\begin{remark}
We have already shown that the oscillations vanish when the grid is refined but the under-resolved simulation still demonstrates the fundamentally different character of the two different regularizations.
\end{remark}

\begin{figure}[ht]
\subfloat[Navier-Stokes]{\includegraphics[width=4.5cm]{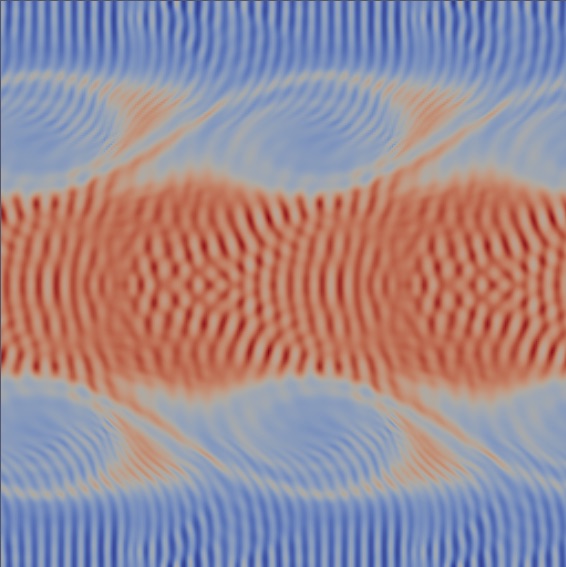}\label{fig44a}}
\subfloat[Eulerian, $\alpha=1$]{\includegraphics[width=4.5cm]{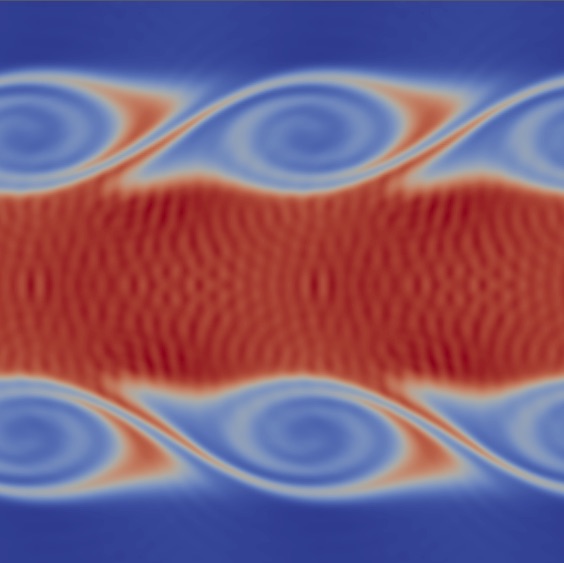}}
\subfloat[Eulerian, $\alpha=4/3$]{\includegraphics[width=4.5cm]{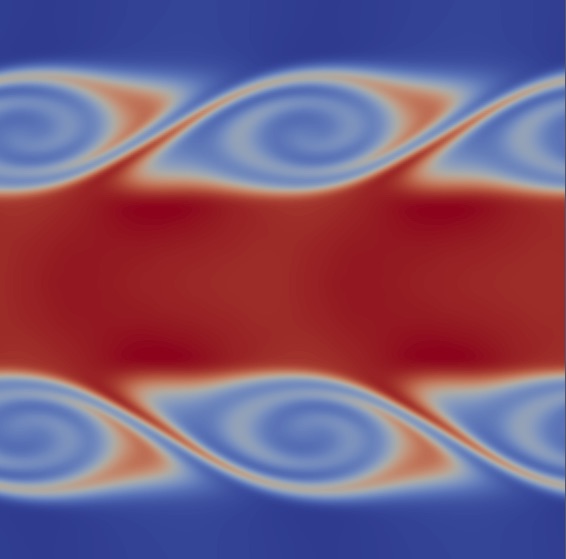}}
\caption{$\rho$ at $T=3$ with $256^2$ grid points. ($\nu=\alpha\mu/\rho$)}
\label{fig44}
\end{figure}

Furthermore, we do not claim that the system (\ref{eulerian}) is immune to oscillations. The system (\ref{eulerian}) has a mode of thermodynamic relaxation but the system is still highly non-linear and might not be diffusive enough in the under-resolved case to control the oscillations fed by the discontinuity. This is seen in the solutions at $T=3$ on the $256^2$ grid that are depicted in  Fig. \ref{fig44}. (Standard colour scheme.) Here, the oscillations from the initial data have grown and are now visible for the $\alpha=1$ solution but the $\alpha=4/3$ is still smooth. However, the Navier-Stokes solution is dramatically different to both the Eulerian solutions.


The oscillatory behaviour is also a source of practical concern. The oscillations can potentially cause completely erroneous solutions, if they trip secondary instabilities that would otherwise not have been tripped. In fact, that is what happens on coarser grids for (\ref{NS}) and (\ref{eulerian}) with $\alpha=1$. Only (\ref{eulerian}) with $\alpha=4/3$ is able to produce the roll-ups on a $128^2$ grid. However, the failure for the two least diffusive simulations is rather an effect of under-resolution than the PDE systems per se.

\subsection{Blast wave}\label{sec:blast}

So far, we have seen that well-resolved numerical solutions are fairly similar for the two systems but the under-resolved simulations indicated the fundamentally different effect of the diffusive terms. 

In this section, we study wall boundary conditions. In Section \ref{sec:adiabatic1}, it was shown that the standard adiabatic wall boundary conditions allow what is usually interpreted as diffusive fluxes of mass, density and entropy through a wall. This anomaly disappears when $\partial_n\rho=0$ is added as a boundary condition for (\ref{eulerian}). However, the addition of this boundary condition has a profound effect on the solution. 

\begin{figure}[ht]
\centering
\subfloat{\includegraphics[width=8cm]{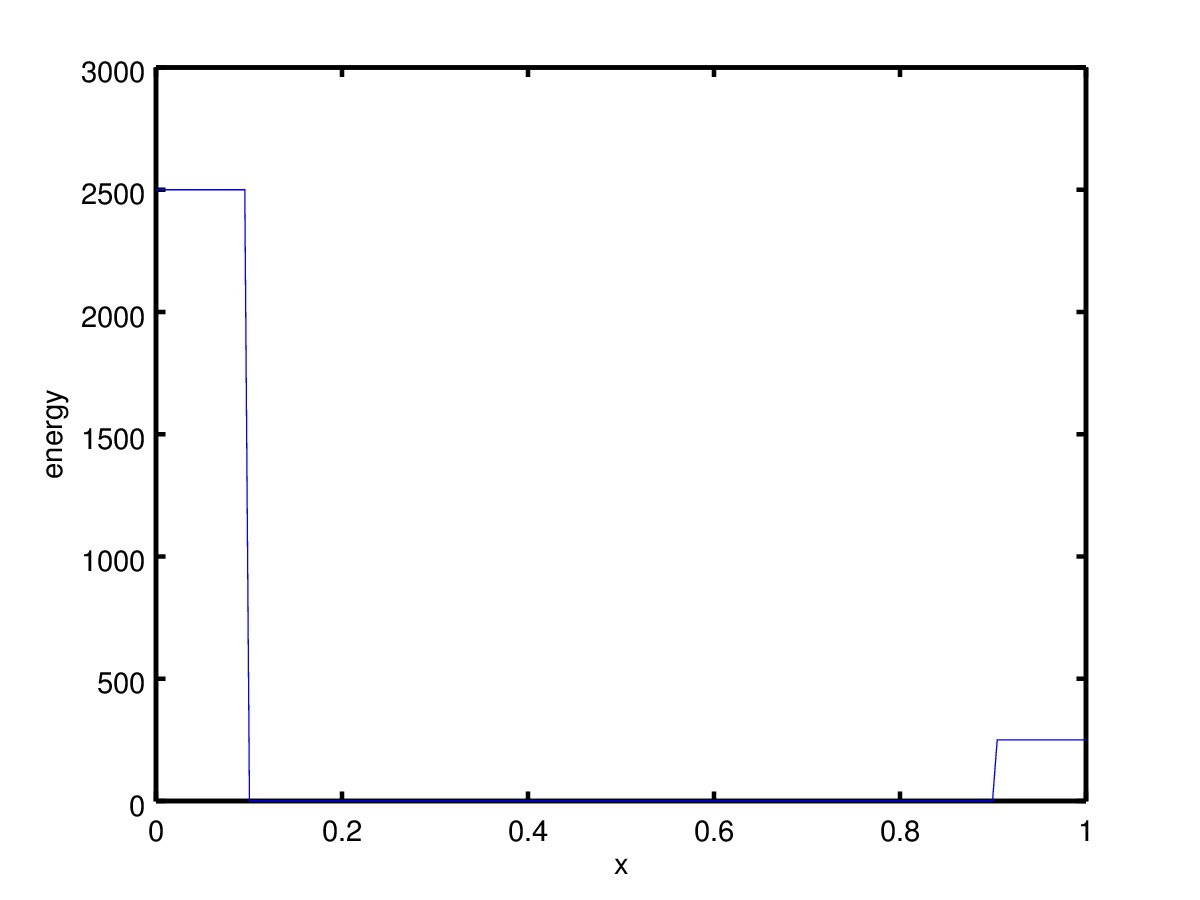}}
\caption{Initial data for the total energy.}
\label{fig_blastInit}
\end{figure}

We have used the following setup for a 1-dimensional blast wave:
\begin{align}
(\rho,u,p)=(1,0,1000), &\quad  0\leq x < 0.1,   \nonumber \\
(\rho,u,p)=(1,0,0.01),\, &\quad 0.1\leq x < 0.9, \nonumber \\
(\rho,u,p)=(1,0,100),\,\, & \quad 0.9\leq x \leq  1.\nonumber 
\end{align}
and 
\begin{align}
Pr=0.72, \quad \gamma=1.4, \quad R=286.84, c_p=1005,\quad \quad \mu=0.1, \quad \kappa=\frac{\mu c_p}{Pr}.\nonumber
\end{align}
The initial condition for the energy is depicted in Fig. \ref{fig_blastInit}.
In this case, the details of the scheme and in particular the implementation of the boundary conditions are essential. The inviscid flux is computed in the same way for both systems and the boundary condition used for the inviscid flux is $u=0$. (Generally, it is only the normal velocity but in 1-D there is only one velocity component.) The grid is $x_i=ih$ where $h$ is the grid size and $i=0...N$ on the domain $(0,1)$. At the interior points, the stencil for the inviscid flux is,
\begin{align}
\partial_x f^j\approx \frac{f^j_{i+1}-f^j_{i-1}}{2h}, 
\end{align}
and at the boundaries 
\begin{align}
\partial_x f^j|_{x=0}\approx \frac{f^j_{1}-f^j_{0}}{h}, \nonumber \\
\partial_x f^j|_{x=1}\approx \frac{f^j_{N}-f^j_{N-1}}{h}, \nonumber 
\end{align}
where $j$ denotes a component of the inviscid flux. The boundary fluxes are obtained by inserting $u=0$.
\begin{align}
f_{0,N}=(0,p_{0,N},0)^T.\label{bflux}
\end{align}
\begin{remark}
The finite difference schemes are based on the Summation-by-parts framework to ensure stable and accurate approximations. (See \cite{SvardNordstrom14}.) The boundary flux (\ref{bflux}) is the entropy consistent (stable) choice. (See \cite{ParsaniCarpenter15, SvardOzcan14}.) 
\end{remark}
The velocity is computed as $v_i=m_i/\rho_i$ and $v_{0,N}=0$ (no-slip boundary conditions). In the viscous fluxes of the Navier-Stokes equations, the following discretizations are used,
\begin{align}
\frac{4\mu}{3}v_{xx}|_{x_i}&\approx \frac{4\mu}{3}\frac{v_{i+1}-2v_i+v_{i-1}}{h^2}, \quad i=1,...,N-1\nonumber \\
\frac{4\mu}{3}v_{xx}|_{x_0}&\approx\frac{4\mu}{3}\frac{v_{2}-2v_1+v_{0}}{h^2}, \nonumber \\ 
\frac{4\mu}{3}v_{xx}|_{x_N}&\approx\frac{4\mu}{3}\frac{v_{N}-2v_{N-1}+v_{N-2}}{h^2}, \nonumber 
\end{align}
(See \cite{MattssonNordstrom04} for a numerical rationale for this approximation.) Furthermore.
\begin{align}
(\frac{4\mu}{3}vv_x)_x|_{x_i}\approx D_+(\frac{4\mu}{3}v_iD_-v_i)\label{ns_vel}
\end{align}
where $D_+w_i=(w_{i+1}-w_i)/h$ and at the right boundary $D_+w_N=(w_{N}-w_{N-1})/h$. Similarly, $D_-w_i=(w_{i}-w_{i-1})/h$ and at the left boundary $D_-w_0=(w_{1}-w_{0})/h$. Finally, enforcing the approximation of $T_x=0$ on the right and left boundaries results in the following scheme for the temperature diffusion,
\begin{align}
\kappa T_x|_{x_i}&\approx\kappa \frac{T_{i+1}-2T_i+T_{i-1}}{h^2}, \quad i=1...N-1\nonumber \\
\kappa T_x|_{x_0}&\approx\kappa \frac{T_{1}-T_0}{h^2}, \nonumber \\
\kappa T_x|_{x_N}&\approx\kappa \frac{-T_{N}+T_{N-1}}{h^2}. \nonumber 
\end{align}
(Once again, we refer to \cite{MattssonNordstrom04} for a rationale.)

Next, we detail the scheme for (\ref{eulerian}). For all three equations we use the following basic scheme for the parabolic flux:
\begin{align}
D_+\nu_iD_-\xi_i,\label{euler_par}
\end{align}
where $D_{+,-}$ are given as above and $\xi$ is either $\rho,m$ or $E$. To enforce the boundary condition for $\rho$, we first compute $dr_i=D_-\rho_i$ at all grid points and set $dr_0=dr_N=0$ which (approximately) enforces the boundary condition $\partial_n\rho=0$.  Next, the viscous flux for $\rho$ is computed as $D_+(\nu_idr_i)$. 

For the momentum equation, we set $m_0=m_N=0$ before we compute (\ref{euler_par}) with $\xi_i=m_i$.

For the energy equation the boundary conditions $\rho_x=T_x=0$ combine to $p_x=0$. Similar to the procedure for the density, we compute the term in three steps. 1) $dE_i=D_-E_i$. Then 2)
\begin{align}
dE_0=\frac{\frac{\rho_1v_1^2}{2}-\frac{\rho_0v_0^2}{2}}{h}\nonumber
\end{align}
where $v_0=0$. (Similarly at $i=N$.) Now that the boundary conditions have been enforced, we proceed with 3) $D_+(\nu dE_i)$.

\begin{figure}[ht]
\centering
\subfloat[Eulerian, $\alpha=1$]{\includegraphics[width=8cm]{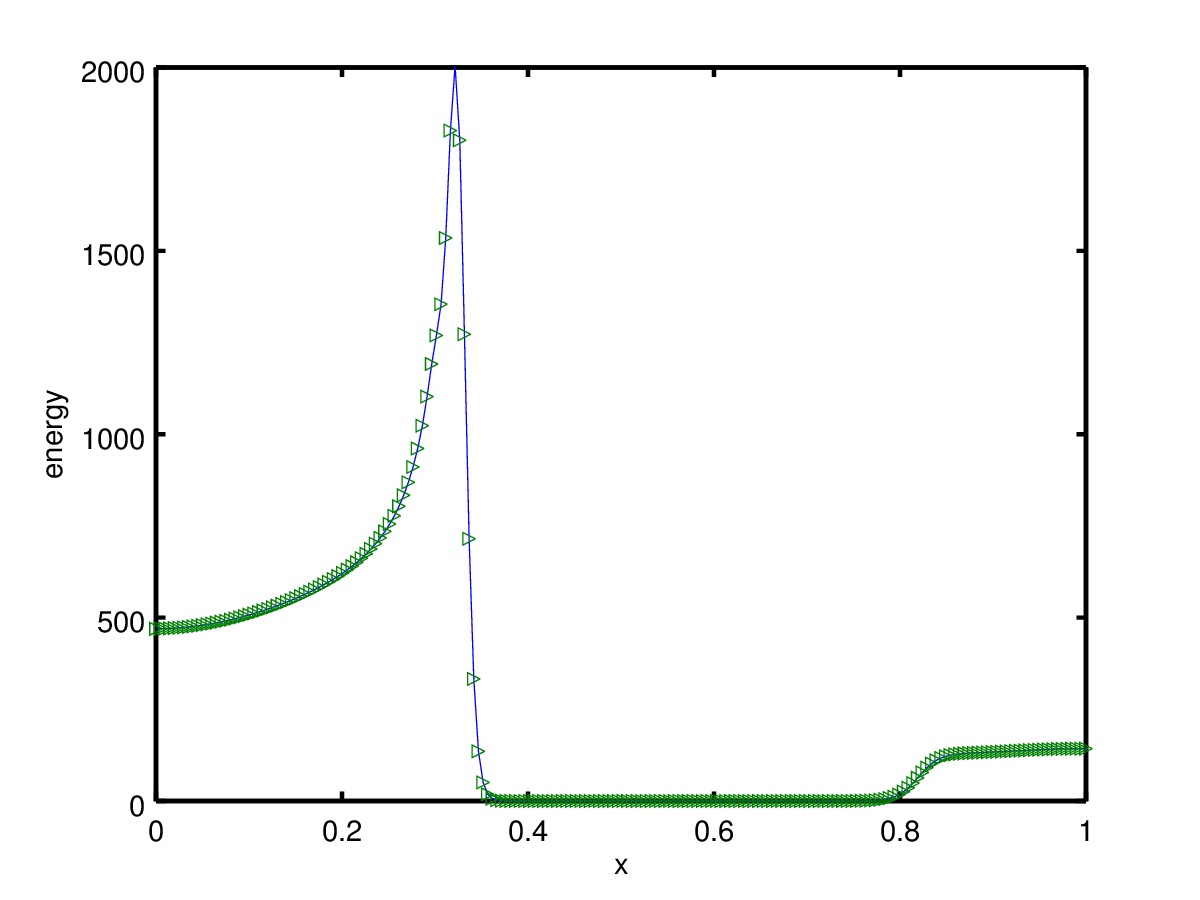}}
\subfloat[Navier-Stokes]{\includegraphics[width=8cm]{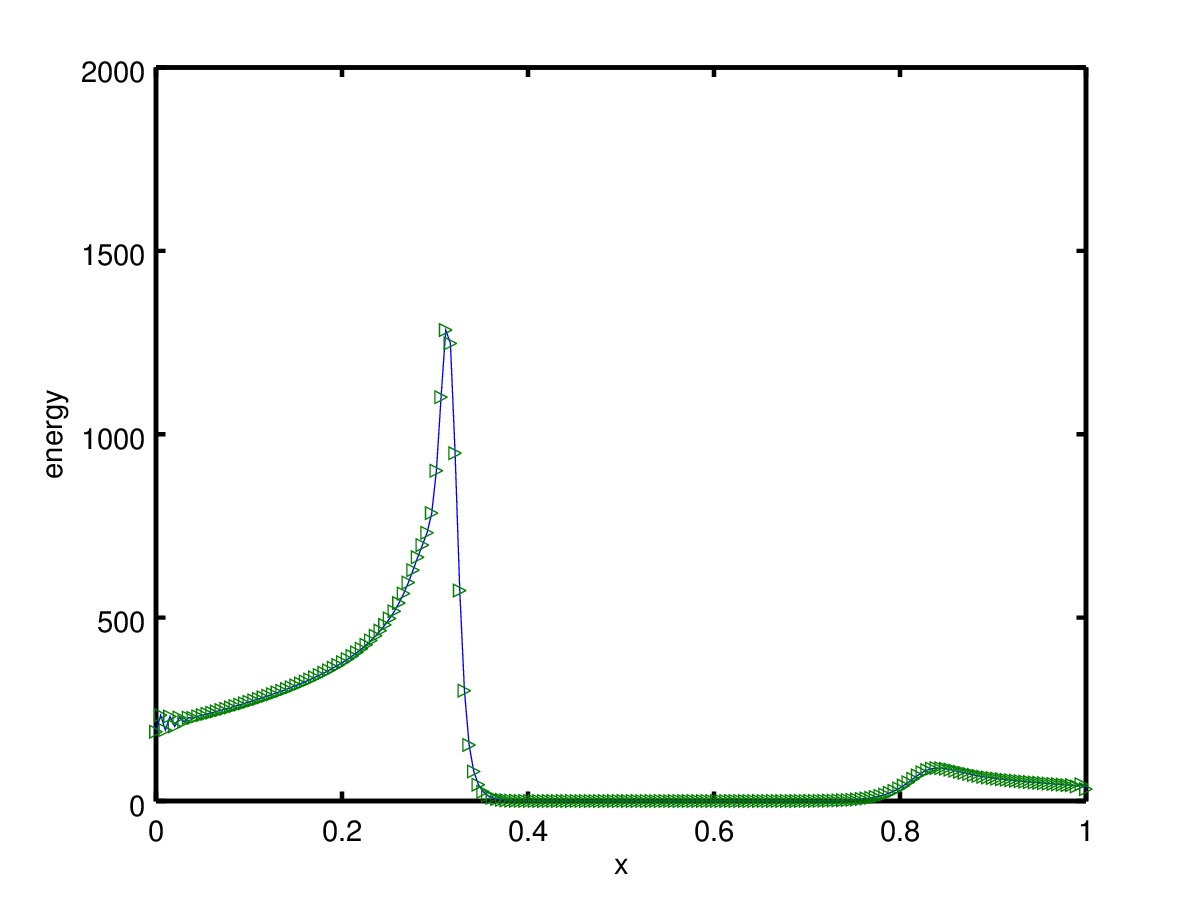}}
\caption{Energy at $t=0.01$. 200 grid points.}
\label{fig_blast_energy}
\end{figure}

\begin{figure}[ht]
\centering
\subfloat[Eulerian, $\alpha=1$]{\includegraphics[width=8cm]{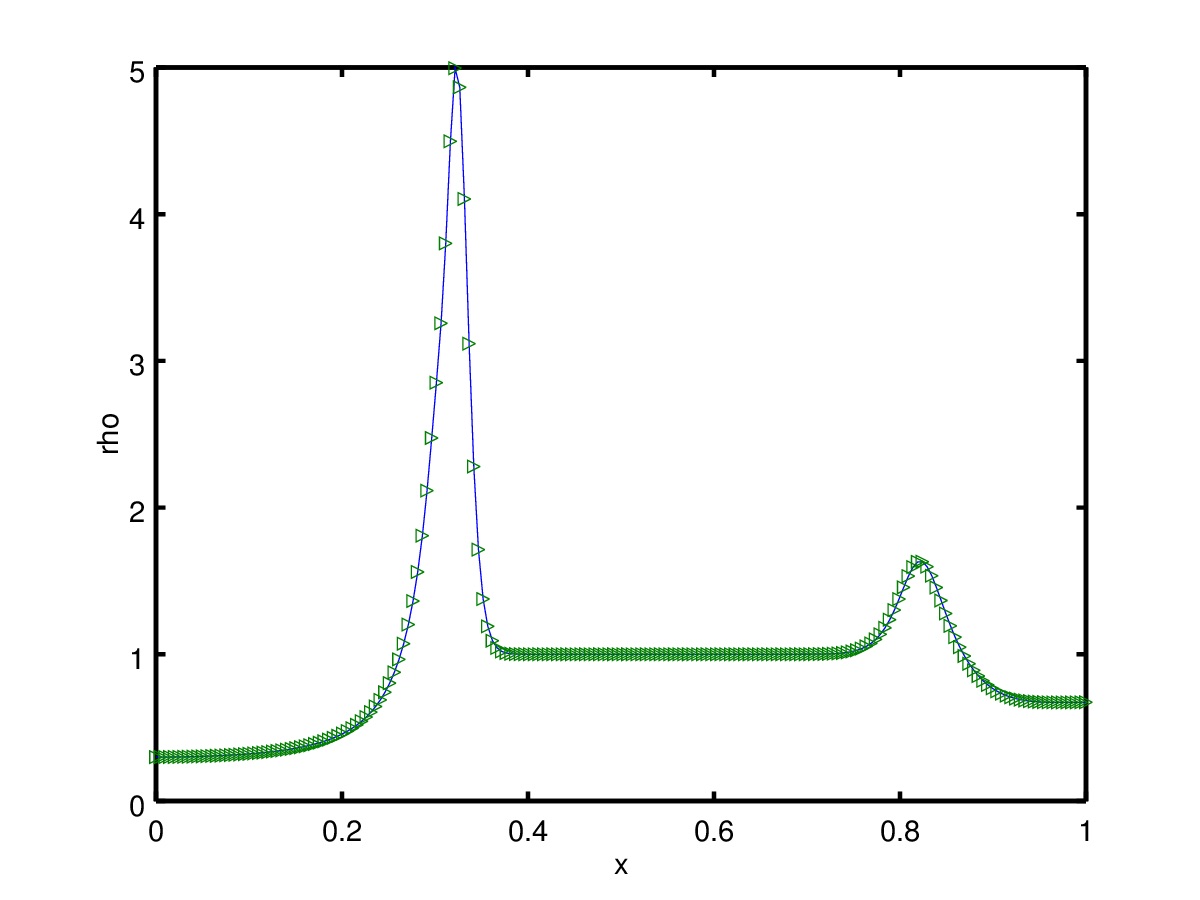}}
\subfloat[Navier-Stokes]{\includegraphics[width=8cm]{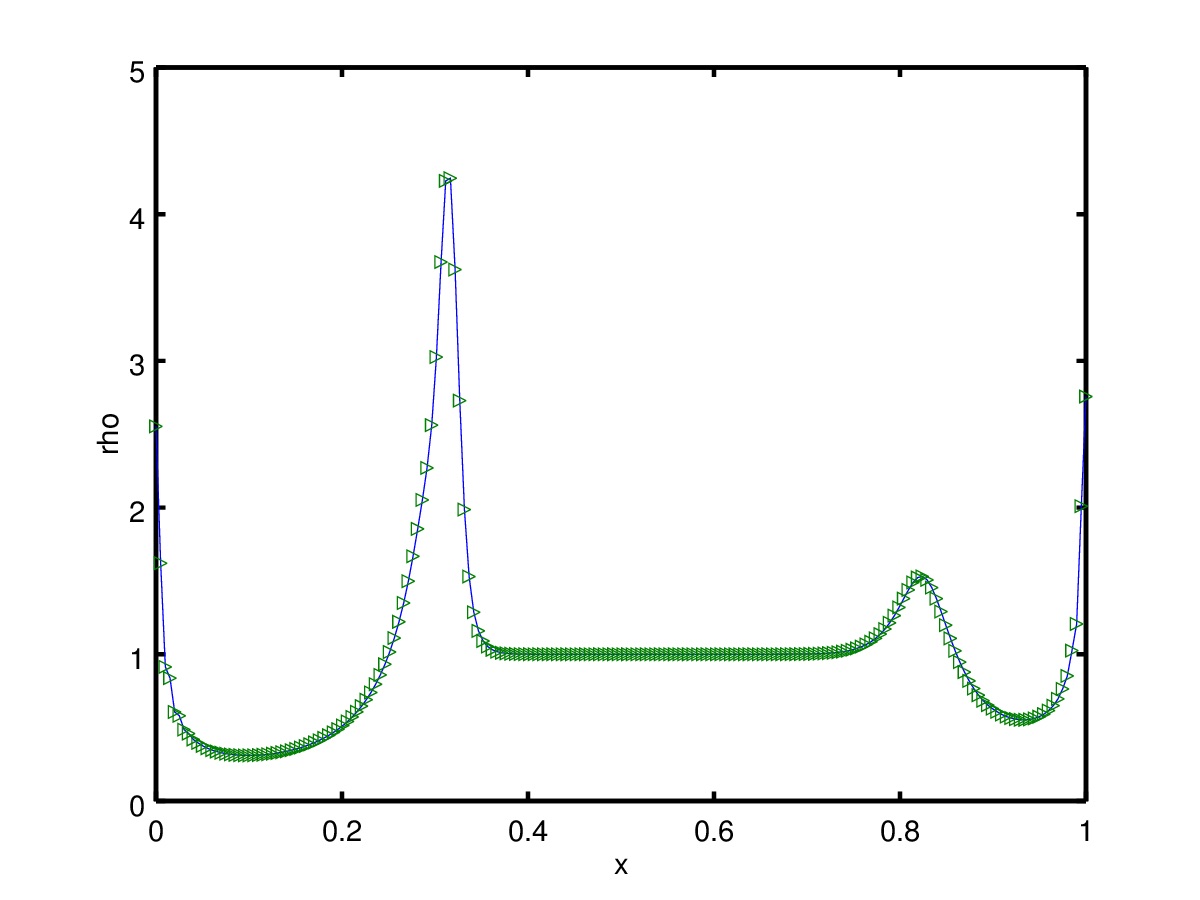}}
\caption{Density at $t=0.01$. 200 grid points.}
\label{fig_blast_density}
\end{figure}

\begin{figure}[ht]
\centering
\subfloat[Eulerian]{\includegraphics[width=8cm]{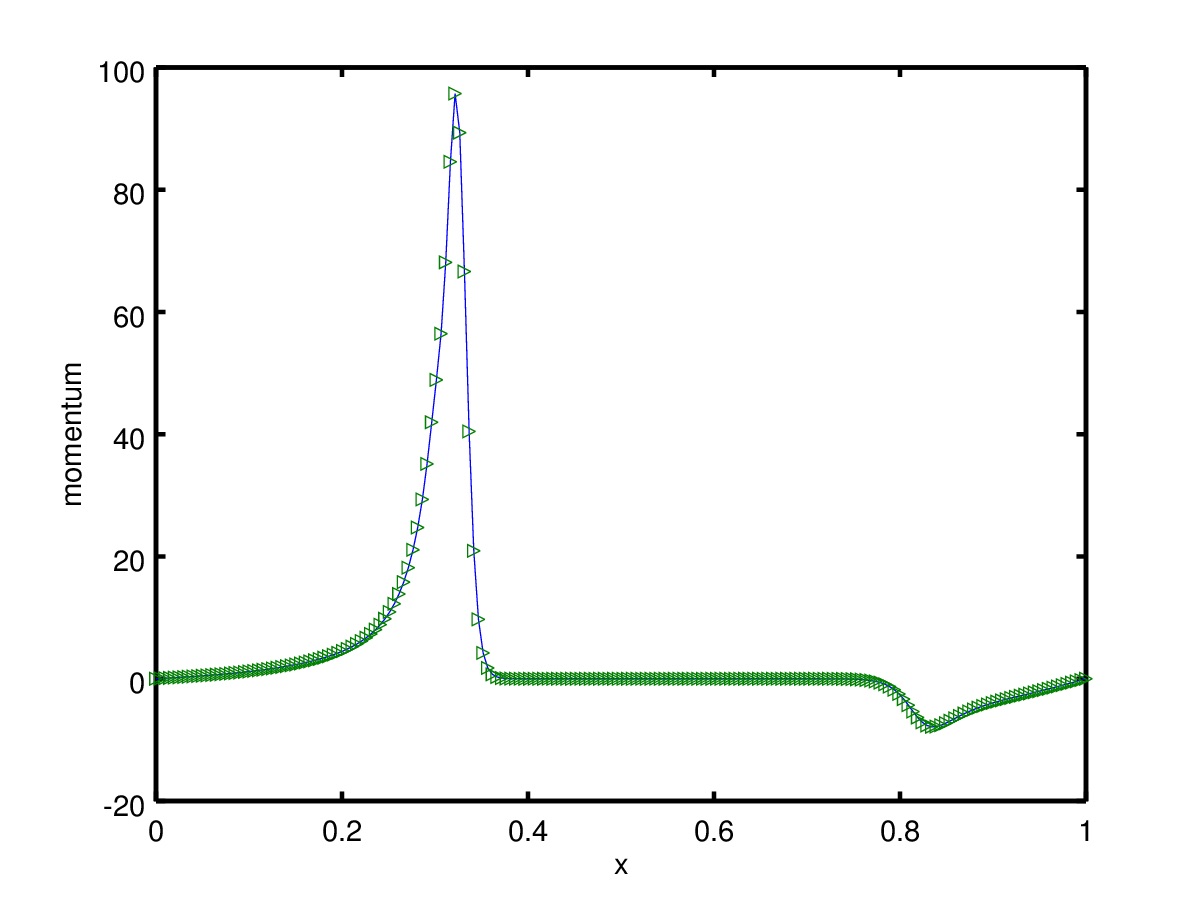}}
\subfloat[Navier-Stokes]{\includegraphics[width=8cm]{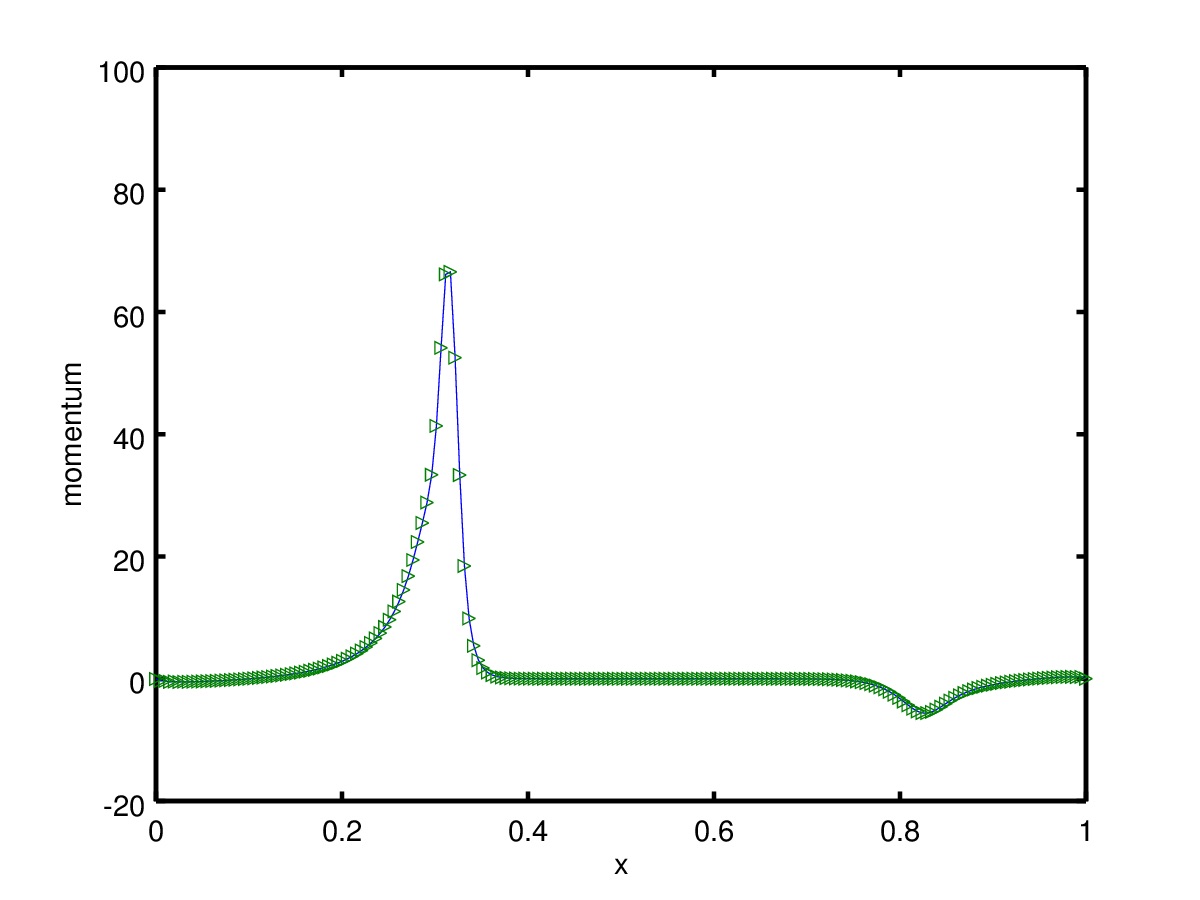}}
\caption{Momentum at $t=0.01$. 200 grid points.}
\label{fig_blast_mom}

\end{figure}

Using these schemes, we have computed the solution at $t=0.01$ for (\ref{NS}) and (\ref{eulerian}). The results are displayed in Fig. \ref{fig_blast_energy}-\ref{fig_blast_mom}. Comparing the energy of the two systems in Fig. \ref{fig_blast_energy}, there are a number of differences. The values at the boundaries are completely different, which in turn affects the strengths of the shocks. Furthermore, there are oscillations in the energy on the left boundary in the Navier-Stokes solution. These oscillations can be damped with a sufficiently strong artificial diffusion but we emphasize that they are not numerical instabilities. Both schemes are numerically stable. The oscillations are simply inaccuracies and are caused by the strong boundary layer in the density present in the Navier-Stokes solution as seen in Fig. \ref{fig_blast_density}. This boundary layer is a consequence of the Navier-Stokes equations not having a boundary condition affecting the density. When the simulation is run, the density is ``lagging'' behind the other variables. (Naturally, the strong gradient is less accurate which affects the energy.) In the Eulerian system it is clear that the boundary condition $\rho_x=0$ is enforced. The difference in the boundary conditions makes the solutions completely different.  

\section{Conclusions}\label{sec:conclusions}

We have presented a number of  physical deficiencies  of the Navier-Stokes equations and demonstrated that their origin is the Lagrangian derivation. By remaining in an Eulerian frame, we deduced the following model for a diffusive and compressible flow:
\begin{align}
\partial_t \rho + div_\xb(\rho \vv )&= \nabla_\xb \cdot (\nu \nabla_\xb \rho)\nonumber \\
\partial_t (\rho \vv) + div_\xb(\rho \vv \otimes \vv) + \nabla_\xb p&=\nabla_\xb \cdot
(\nu \nabla_\xb \rho \vv)) \tag{\ref{eulerian}} \\
\partial_t (E) + div_\xb(E \vv +p\vv)  &= \nabla_\xb \cdot( \nu \nabla_\xb E) \nonumber \\ 
p&=\rho R T \quad \textrm{ideal gas law}\nonumber  
\end{align}
where $\nu=\alpha\mu/\rho$ is proportional to the kinematic viscosity. We propose that (\ref{eulerian}) replaces the standard Navier-Stokes system (\ref{NS}).


\begin{acknowledgments}
The author would like to thank Dr. Mark H. Carpenter and Professor Jarle Berntsen for valuable discussions. 
\end{acknowledgments}

\appendix

\section{Galilean invariance of the energy equation}\label{app:galilean}
The energy equation is,
\begin{align}
E_t+ (u(E+p))_x +(v(E+p))_y=(\nu E_{x})_x+(\nu E_{y})_y.\nonumber 
\end{align}
We use $E=\frac{1}{2}\rho(u^2+v^2)+\frac{p}{\gamma-1}$ and apply the transformation,
\begin{align}
(\frac{1}{2}\rho((U+u')^2+(v')^2)+\frac{p}{\gamma-1})_\tau
-U(\frac{1}{2}\rho((U+u')^2+(v')^2)+\frac{p}{\gamma-1})_\xi\nonumber \\
+ ((U+u')(\frac{1}{2}\rho((U+u')^2+(v')^2)
+\frac{p}{\gamma-1}+p))_\xi \nonumber \\
+(v'(\frac{1}{2}\rho((U+u')^2+(v')^2)+\frac{p}{\gamma-1}+p))_\eta&=\nonumber\\
(\nu(\frac{1}{2}\rho((U+u')^2+(v')^2)+\frac{p}{\gamma-1})_{\xi})_\xi+(\nu (\frac{1}{2}\rho((U+u')^2+(v')^2)+\frac{p}{\gamma-1})_{\eta})_\eta.\nonumber 
\end{align}
Use $E'=\frac{1}{2}\rho((u')^2+(v')^2)+\frac{p}{\gamma-1}$, and recast the equation
\begin{align}
E'_{\tau}+
(\frac{1}{2}\rho(U^2+2Uu'))_\tau
-UE'_{\xi}
-U(\frac{1}{2}\rho(U^2+2Uu')_\xi\nonumber \\
(u'(E'+p))_{\xi}
+ (U(\frac{1}{2}\rho((U+u')^2+(v')^2)
+\frac{p}{\gamma-1}+p))_\xi 
+ (u'(\frac{1}{2}\rho(U^2+2Uu'))_{\xi}\nonumber \\
+(v'(E'+p)_\eta
+(v'(\frac{1}{2}\rho(U^2+2Uu')))_\eta
&=\nonumber\\
(\nu E'_{\xi})_\xi+(\nu E'_{\eta})_\eta+
(\nu(\frac{1}{2}\rho((U^2+2Uu'))_{\xi})_\xi+(\nu(\frac{1}{2}\rho((U^2+2Uu'))_{\eta})_\eta.\nonumber 
\end{align}
Rewrite
\begin{align}
E'_{\tau}+
(\frac{1}{2}\rho(U^2+2Uu'))_\tau
-UE'_{\xi}
-U(\frac{1}{2}\rho(U^2+2Uu')_\xi\nonumber \\
(u'(E'+p))_{\xi}
+ (U(E'+\frac{1}{2}\rho((U^2+2Uu')+p)_\xi 
+ (u'(\frac{1}{2}\rho(U^2+2Uu'))_{\xi}\nonumber \\
+(v'(E'+p)_\eta
+(v'\frac{1}{2}\rho(U^2+2Uu'))_\eta
&=\nonumber\\
(\nu E'_{\xi})_\xi+(\eta E'_{\eta})_\eta+
(\nu (\frac{1}{2}\rho((U^2+2Uu'))_{\xi})_\xi+(\nu (\frac{1}{2}\rho((U^2+2Uu'))_{\eta})_\eta.\nonumber 
\end{align}
A number of terms cancel:
\begin{align}
E'_{\tau}+
(\frac{1}{2}\rho(U^2+2Uu'))_\tau+
(u'(E'+p))_{\xi}
+ (u'(\frac{1}{2}\rho(U^2+2Uu'))_{\xi}\nonumber \\
Up_{\xi}+(v'(E'+p)_\eta
+(v'\frac{1}{2}\rho(U^2+2Uu'))_\eta
&=\nonumber\\
(\nu E'_{\xi})_x+(\nu E'_{\eta})_\eta+
(\nu (\frac{1}{2}\rho(U^2+2Uu'))_{\xi})_\xi+(\nu (\frac{1}{2}\rho(U^2+2Uu'))_{\eta})_\eta.&\nonumber 
\end{align}
The terms multiplied by $\frac{1}{2}U^2$ combine to the continuity equation
and cancel,
\begin{align}
E'_{\tau}+
(\frac{1}{2}\rho(2Uu'))_\tau+
(u'(E'+p))_{\xi}
+ (u'(\frac{1}{2}\rho(2Uu'))_{\xi}\nonumber \\
Up_{\xi}+(v'(E'+p)_\eta
+(v'\frac{1}{2}\rho(2Uu'))_\eta
&=\nonumber\\
(\nu E'_{\xi})_\xi+(\nu E'_{\eta})_\eta+
 (\nu (\frac{1}{2}\rho(2Uu'))_{\xi})_\xi+(\nu (\frac{1}{2}\rho((2Uu'))_{\eta})_\eta.\nonumber 
\end{align}
The terms multiplied by $U$ are cancelled by the x-momentum equations and we are left with
\begin{align}
E'_{\tau}+(u'(E'+p))_{\xi}
+(v'(E'+p))_\eta
&=
(\nu E'_{\xi})_\xi+(\nu E'_{\eta})_\eta.\nonumber
\end{align}
Hence, the energy equation is Galilean invariant, which concludes the proof.

 \section{Booster density}\label{sec:Booster}

 For a collection of particles a uniform centre-of-mass movement is expected when there are no external forces.  For an Eulerian continuum model this property can not be expected since it would require knowledge of the centre-of-mass within each fluid volume. Nevertheless, under certain conditions, the system (\ref{eulerian}) has this property. 

If the booster density (vector) $b_i=\rho x_i - (\rho u_i)t$ is conserved locally, i.e., satisfies a conservation law, the centre-of-mass movement is uniform. We derive an equation for the evolution of $b_i$. We carry out the derivations in two space dimensions to reduce notation. First and by definition,
\begin{align}
\partial_t b_1= x \partial_t  \rho  -(\rho u) - (\rho u)_t t.\nonumber
\end{align}
 Using the equations for conservation of mass and momentum, we arrive at,
\begin{align}
\partial_t b_1=  (-(x\rho u)_x -(x \rho v)_y +(\nu x \rho_x)_{x}+(\nu x\rho_{y})_y) -\nu \rho_x \nonumber \\
- (-(t(\rho u^2+p))_x-(t\rho uv)_y+(t\nu(\rho u)_x)_{x}+(t\nu(\rho u)_y)_{y}).\nonumber
\end{align}

Similarly for $b_2$:
\begin{align}
\partial_t b_2=  (-(y\rho u)_x -(y\rho v)_y +(\nu y \rho_{x})_x+(\nu y\rho_{y})_y)-\nu \rho_y   \nonumber \\
- (-(t\rho u v)_x-(t(\rho v^2+p))_y+((t\nu(\rho v)_x)_{x})+(t\nu(\rho v)_y)_{y})\nonumber
\end{align}

The equations for $b_1$ and $b_2$ are conservative, if $\nu \rho_x$ and $\nu\rho_y$ are conservative terms. This is the case if $\nu=\nu(\rho)$ and integrable such that there is a $N'(\rho)=\nu(\rho)$. However, the diffusion coefficients are generally functions of temperature, so this limitation is not very useful. Finally, we stress once again that there is no reason to enforce this property on an Eulerian continuum model as it is not satisfied for averages of particles inside a fluid volume.

\section{Angular momentum}\label{sec:ang_mom}

This Section contains a proof that the macroscopic angular momentum is conserved when $\nu=constant$.

The angular momentum is given as,  $\lbf=\xb\times \rho \uu$. Consequently, a conservation law should obtained by crossing $\xb$ with the momentum equations, since 
$\xb\times (\rho \uu)_t=(\xb\times \rho \uu)_t=\lbf_t$. In \cite{Ottinger09}, it
was shown that for the Navier-Stokes system, a system conservation laws is
obtained. Hence, we can immediately conclude that the inviscid Euler equations satisfies an angular momentum  equation of the form
\begin{align}
\frac{\partial \lbf}{\partial t}+div_x {{\bf \bar F^{Euler}}}=0.\nonumber
\end{align}
Hence, we only need to check if the diffusion of (\ref{eulerian}) also turns into divergence form for the angular momentum. Using the notation $\Fv^m=\nu \partial_x \uu$, we have

\begin{align}
\xb\times \partial_x\left(\begin{array}{c} F^{m2}\\F^{m3}\\F^{m4} \end{array}\right)=
\left(\begin{array}{c} 
yF^{m4}_x-zF^{m3}_x \\
-xF^{m4}_x+zF^{m2}_x \\
xF^{m3}_x-yF^{m2}_x 
\end{array}\right)&=
\left(\begin{array}{c} 
(yF^{m4})_x-(zF^{m3})_x \\
-(xF^{m4})_x+F^{m4}+(zF^{m2})_x \\
(xF^{m3})_x-F^{m3}-(yF^{m2})_x 
\end{array}\right).\nonumber
\end{align}
The contributions from the other two fluxes, are given as
\begin{align}
\xb\times \partial_y\Gv^l&=
\left(\begin{array}{c} 
(yG^{m4})_y-G^{m4}-(zG^{m3})_y \\
(-xG^{m4}+zG^{m2})_y \\
(xG^{m3}-yG^{m2})_y+G^{m2} 
\end{array}\right), \nonumber \\
\xb \times \partial_z\Hv^l &=
\left(\begin{array}{c} 
(yH^{m4}-zH^{m3})_z+H^{m3} \\
(-xH^{m4}+zH^{m2})_z-H^{m2} \\
(xH^{m3}-yH^{m2})_z
\end{array}\right).\nonumber 
\end{align}
Most of the terms in these vectors are conservative but there are a few that
may not be:
\begin{align}
H^{m3}-G^{m4},\quad
F^{m4}-H^{m2},\quad
G^{m2}-F^{m3}. \nonumber  
\end{align}
However, we note that the macroscopic angular momentum would be conserved if all
individual components above of were partial derivatives. That is only the case if $\nu=constant$.

\section{The numerical scheme}\label{app:scheme}

For a scalar quantity $y_i=y(x_i)$ where $i$ is the grid-position index, we define
\begin{align}
D_0y_i&=\frac{y_{i+1}-y_{i-1}}{2h} \approx y_x(x_i)\nonumber\\
D_+y_i&=\frac{y_{i+1}-y_{i}}{2h} \approx y_x(x_i)\nonumber\\
D_-y_i&=\frac{y_{i}-y_{i-1}}{2h} \approx y_x(x_i)\nonumber\\
D_2y_i&=D_+D_-y_i=\frac{y_{i+1}-2y_i+y_{i-1}}{h^2} \approx y_{xx}(x_i) \nonumber
\end{align}
where is the grid size $h=x_{i+1}-x_i$. The (semi-discrete) scheme for the one-dimensional version of (\ref{NS}) at the interior points is:
\begin{align}
(\rho_i)_t + D_0m_i&=0\nonumber \\
(m_i)_t + D_0(u_im_i+p_i)&=\frac{4\mu}{3} D_2u_i\nonumber \\
(E_i)_t + D_0(u_i(E_i+p_i))&=\frac{4\mu}{3} D_0(u_iD_0u_i)+\kappa D_2T_i\nonumber 
\end{align}
In the scheme for (\ref{eulerian}) the right-hand side is replaced with 
\begin{align}
D_0\nu_iD_0 u_i\nonumber
\end{align}
where $u_i$ is either $\rho_i,m_i$ or $E_i$. At the boundaries the schemes changed to skew stencils. (See \cite{SvardNordstrom14}.)  The scheme is marched in time with a third-order strong-stability preserving Runge-Kutta scheme.

The 2-D scheme used for the Kelvin-Helmholtz problem is a  straightforward generalization of the above scheme. The only differences are that the boundary conditions are periodic and the stress tensor is computed with the $D_{\pm}$ approximations. Similarly, for the Eulerian system the diffusive approximations are produced by $D_-\nu_iD_+u_i$. 

These stencils ensure that there is no odd-even decoupling which otherwise may occur in periodic problems. Hence, the highest mode is always affected by the parabolic damping. The oscillations occurring in the under-resolved Navier-Stokes simulation (see Fig. \ref{fig3}) are therefore a product of the equations and not the discretization.

\bibliographystyle{plainnat}
\bibliography{ref}

\end{document}